\documentclass[a4paper, 12pt]{article}
\usepackage{mlmodern}
\usepackage{microtype}
\usepackage{amsmath, amssymb}
\usepackage{bm}
\usepackage[dvipsnames]{xcolor}
\usepackage{tikz}
\usetikzlibrary{calc, arrows.meta, decorations.markings, bending}
\usepackage{titling}
\usepackage{authblk}
\usepackage{cite}

\usepackage[
    colorlinks,
    citecolor = Green,
    linkcolor = blue,
    urlcolor  = blue
]{hyperref}

\usepackage[
    left   = 2cm,
    right  = 2cm,
    top    = 2.5cm,
    bottom = 2cm
]{geometry}

\begin{document}
\title{Orbital antiferromagnetic currents \\ in a frustrated fermionic ladder}

\author[1,2,\thanks{\href{mailto:shota_garuchava@hotmail.com}{shota\_garuchava@hotmail.com}}]{Shota Garuchava}

\author[1,2,\thanks{\href{mailto:giajaparidze@gmail.com}{giajaparidze@gmail.com}}]{G.I. Japaridze}

\author[1,2,3,\thanks{\href{mailto:alex.a.nersesyan@gmail.com}{alex.a.nersesyan@gmail.com} (corresponding author)}]{A.A. Nersesyan}

\affil[1]{\small\textit{Ilia State University, 0162 Tbilisi, Georgia}}

\affil[2]{\small\textit{The Andronikashvili Institute of Physics, 0177 Tbilisi, Georgia}}

\affil[3]{\small\textit{The Abdus Salam International Centre for Theoretical Physics, 34151, Trieste, Italy}}

\date{}

\maketitle

\renewcommand{\abstractname}{\vspace{-4\baselineskip}}
\begin{abstract}

We consider a spinless  $t$-$t'$ ionic Hubbard chain at 1/2 filling and large hopping ratio $t'/t$. In this limit
the model adequately maps onto  a weakly coupled  triangular ladder with a potential interchain bias. The low-energy
properties of the system are formed due to the interplay of geometrical frustration, correlations and charge imbalance.
We derive the effective field-theoretical model to study universal properties of the model in the scaling limit. We show
that at full  dynamical frustration the ground state of the ladder represents a repulsive version of the Luther-Emery liquid
with dominant orbital antiferromagnetic correlations exhibiting the slowest power-law decay in the ground state.
Pairing correlations also display algebraic order but are subdominant.
At an incomplete dynamical frustration  a finite commensurability gap is dynamically generated, and the fluctuating OAF 
transforms to a long-range ordered state with a spontaneously broken time reversal symmetry. The mass gap in the spectrum
of relative density fluctuations gets suppressed upon increasing the potential bias.

\end{abstract}

\section{Introduction\label{sec:Introduction}}
The problem of structural distortions or external symmetry breaking fields in strongly correlated electron systems has been of much interest in recent decades. In the context of quasi-one-dimensional (1D) organic materials, the 1D repulsive 1/2-filled Hubbard model with a staggered scalar potential, known as the ionic Hubbard model (IHM), has played a paradigmatic role~\cite{nagaosa1,nagaosa2,nagaosa3,egami}. On the basis of a field-theoretical approach, the ground state phase diagram of the IHM was derived~\cite{fgn1,fgn2}, showing that the Mott insulator and band insulator phases are separated by an intermediate, spontaneously dimerized phase. The two-transition scenario was confirmed in subsequent analytical and numerical works~\cite{torio,manmana,zhang,tincani}. In more recent studies  the interplay between ionicity and correlations has been addressed under different contexts, such as a mass imbalanced Hubbard chain~\cite{sekania}, level statistics~\cite{marco}, inclusion of density-dependent nearest-neighbor (n.n.)\ hopping~\cite{hallberg}, as well as a next-nearest-neighbor (n.n.n.)\ single-particle hopping~\cite{rossini}, IHM with higher symmetry groups~\cite{shan}, spin-dependent staggered potentials~\cite{jn-19}, etc. The studies of ionicity in correlated systems have also been extended to ladder models of interacting fermions~\cite{cnn-1, cnn-2}.

Current interest in the IHM  is enhanced by its recent realization in systems of ultracold atoms on optical lattices~\cite{tarruell,jotzu,messer,loida}. In particular, these objects represent an excellent testing ground for adding to the interplay of correlations and symmetry breaking potentials one more element -- geometrical lattice frustration. In 1D systems, particle-hole symmetry-breaking kinetic frustration  emerges when a n.n.n.\ hopping ($t'$) is added to the n.n.\ one ($t$). Frustration enhances quantum fluctuations of physical quantities and leads to new patterns of correlations in the system at low energies.

%%%%%%%%%%%%%%%%%%%%%%%%%%%%%%%%%%%%%%%%%%%%%%%%%%%%%%%%%%%%%%%%%%%%%%%%%%%%%%%%%%%%%%%%%%%%%%%%%%%%
%%%%%%%%%%%%%%%          Fig 1 
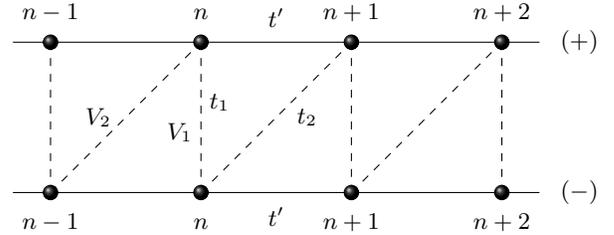
\begin{figure}[t!]
\centering
\begin{tikzpicture}
%%%%%%%%%%%%%%%%%%%%%%%%%%%%%%%%%%%%%%%%%%%%%%%%%%
% definitions
\newcommand{\nSegments}{3}% = nPoints - 1
\newcommand{\pointDistance}{2}% distance between two points on the same subchain
\newcommand{\xOffset}{0.5}% horizontal offset beyond farthest points
\newcommand{\xmin}{-\xOffset}
\pgfmathsetmacro{\xmax}{\pointDistance*\nSegments + \xOffset}
\newcommand{\y}[1]{% y-coordinates of the two subchains
	\ifnum#1=0 0\fi%
	\ifnum#1=1 2\fi%
}
\newcommand{\point}[2]{({#1 * \pointDistance}, \y{#2})}% (x,y) pair of each point
\newcommand{\pointColor}[1]{%
	\ifnum#1=0 red\fi%
	\ifnum#1=1 blue\fi%
}
\newcommand{\pointSize}{3pt}
\newcommand{\indices}[1]{% labels of the indicated points
	\ifnum#1=1 $n-1$\fi%
	\ifnum#1=2 $n\vphantom{n+-1}$\fi%
	\ifnum#1=3 $n+1$\fi%
	\ifnum#1=4 $n+2$\fi%
}
%%%%%%%%%%%%%%%%%%%%%%%%%%%%%%%%%%%%%%%%%%%%%%%%%%
% straight lines
\foreach \i in {0,1}
	\draw (\xmin, \y{\i}) -- (\xmax, \y{\i});
%%%%%%%%%%%%%%%%%%%%%%%%%%%%%%%%%%%%%%%%%%%%%%%%%%
% zig-zag lines
\foreach \i in {0,...,\nSegments}
	\draw[dashed] \point{\i}{1} -- \point{\i}{0};
\foreach \i in {0,...,\number\numexpr \nSegments-1 \relax}
	\draw[dashed] \point{\i}{0} -- \point{\number\numexpr \i+1 \relax}{1};
%%%%%%%%%%%%%%%%%%%%%%%%%%%%%%%%%%%%%%%%%%%%%%%%%%
% points 
\foreach \j in {0,1}
	\foreach \i in {0,...,\nSegments}
		\shade[ball color=black] \point{\i}{\j} circle (\pointSize);
%%%%%%%%%%%%%%%%%%%%%%%%%%%%%%%%%%%%%%%%%%%%%%%%%%
% indicated indices
\foreach \i in {1,2,3,4} {
	\node[below=0.5em] at \point{\number\numexpr \i-1 \relax}{0} {\indices{\i}};
	\node[above=0.5em] at \point{\number\numexpr \i-1 \relax}{1} {\indices{\i}};
}
%%%%%%%%%%%%%%%%%%%%%%%%%
\node[right=0.5em] at (\xmax, \y{1}) {$(+)$};
\node[right=0.5em] at (\xmax, \y{0}) {$(-)$};
%%%%%%%%%%%%%%%%%%%%%%%%%%%%%%%%%%%%%%%%%%%%%%%%%%
% parameters
\node[above=0.25em] at (1.5*\pointDistance, \y{1}) {$t'$};
\node[below=0.25em] at (1.5*\pointDistance, \y{0}) {$t'$};
\node[right] at (\pointDistance, 0.6*\y{1}) {$t_{1}$};
\node[right=0.5em] at (1.5*\pointDistance, 0.5*\y{1}) {$t_{2}$};
\node[left] at (\pointDistance, 0.4*\y{1}) {$V_{1}$};
\node[left=0.25em] at (0.5*\pointDistance, 0.5*\y{1}) {$V_{2}$};
%%%%%%%%%%%%%%%%%%%%%%%%%%%%%%%%%%%%%%%%%%%%%%%%%%
\end{tikzpicture}
\caption{Triangular ladder displayed as a rectangular ladder with one diagonal link added.}
\label{fig:zigzag}
\end{figure}
%%%%%%%%%%%%%%%%%%%%%%%%%%%%%%%%%%%%%%%%%%%%%%%%%%%%%%%%%%%%%%%%%%%%%%%%%%%%%%%%%%%%%%%%%%%%%%%%%%%%

In the recent study of the  $t$-$t'$ extended IHM~\cite{rossini} it has been shown that for a certain range of its parameters the spontaneously dimerized gapped phase of the standard IHM~\cite{fgn1,fgn2} is washed out by frustration  
and replaced by a metallic phase characterized by coexisting charge-density (CDW) and bond-density-wave (BDW) 
correlations. The analysis of Ref.~\cite{rossini} concerns the case ${t' < t}$ in which the standard single-chain 
IHM serves as a zero-order approximation. It is therefore natural to extend consideration to the opposite limit 
${t' \gg t}$ where a more adequate representation of the model is a two-chain ladder with a triangular geometry of the plaquettes, as shown in Fig.~\ref{fig:zigzag}. In this representation, the meaning of the setup parameters is completely different: $t'$ becomes  the in-chain hopping amplitude, the degree of kinetic  frustration is measured by the ratio 
of the alternating  interchain hopping amplitudes $t_1$ and $t_2$, the dynamical frustration is determined by the relationship between the interchain interaction constants $V_1$ and $V_2$, whereas the staggered potential of the 
$t$-$t'$ chain transforms to a spatially uniform interchain potential supporting the relative charge imbalance in the 
triangular ladder system. So the emerging problem becomes interesting in its own right.

In this paper we focus on the spinless version of the 1/2-filled $t$-$t'$ IHM in the large-$(t'/t)$ limit. This brings 
us to the model of tight-binding fermions on a triangular ladder at 1/2-filling, which includes repulsive interactions across the zigzag links and charge imbalance on the two chains. The Hamiltonian of the model reads 
%%%%%%%%%%%%%%%%%%%%%%%%%%%%%%%%%%%%%%%%%%%%%%%%%%%%%%%%%%%
%%%%%%%%%%%     Eq. 1
%%%%%%%%%%%%%%%%%%%%%%%%%%%%%%%%%%%%%%%%%%%%%%%%%%%%%%%%%%%%%%%%%%%%%%%%%%%%%%%%%%%%%%%%%%%%%%%%%%%%
\begin{align}
\label{ladder}
	H
	=
	&-
	t_0
	\sum_{n\sigma}
		\left(
			c^{\dag}_{n\sigma} c^{\vphantom{\dag}}_{n+1,\sigma}
			+
			\mathrm{H.c.}
		\right)
    -
	\sum_{n}
		\left(
			t_1 c^{\dag}_{n+} c^{\vphantom{\dag}}_{n-}
			+
			t_2 c^{\dag}_{n+1,+} c^{\vphantom{\dag}}_{n -} 
			+
			\mathrm{H.c.}
		\right)
\nonumber
\\
	&
	-
	\Delta
	\sum_{n\sigma}
		\sigma
		\hat{n}_{n\sigma}
    +
	\sum_n
		\left(
			V_1 \hat{n}_{n+} \hat{n}_{n-}
			+
			V_2 \hat{n}_{n+} \hat{n}_{n-1,-}
		\right)
.
\end{align}
%%%%%%%%%%%%%%%%%%%%%%%%%%%%%%%%%%%%%%%%%%%%%%%%%%%%%%%%%%%%%%%%%%%%%%%%%%%%%%%%%%%%%%%%%%%%%%%%%%%%
Here the index ${\sigma = \pm}$ labels the chains, the lattice sites $n$ take values ${1,2,\ldots,N}$,
$c^{\dag}_{n\sigma}$ and $c_{n\sigma}$ are the creation and annihilation operators of fermions on the site $n$ 
of chain $\sigma$, and ${\hat{n}_{n\sigma} =  c^{\dag}_{n\sigma} c_{n\sigma}}$ are the occupation number operators.
The in-chain hopping amplitude $t_0$ replaces the n.n.n.\ hopping parameter $t'$ of the extended IHM.
We distinguish between the interchain hopping amplitudes $t_1$ and $t_2$ related to the two neighboring 
zigzag links of the ladder, and we also parametrize two-particle interaction on these links by the constants  
$V_1$ and $V_2$. The $\Delta$-term in \eqref{ladder} describes the applied interchain potential bias.
In the analysis that follows we find it  convenient to treat $\sigma$ as an effective spin-1/2 variable. Then the 
ladder model~\eqref{ladder}, originally defined for spinless fermions, translates into a purely 1D model of 
spin-1/2 interacting fermions with local spin-flip processes ($t_1, t_2$) and a ``magnetic field'' $\Delta$ 
along the $z$-axis of ``spin'' space. In what follows, the total particle density will be dubbed the ``charge'', 
whereas the relative  degrees of freedom  the ``spin''.

We will be concerned with the weak-coupling case:
%%%%%%%%%%%%%%%%%%%%%%%%%%%%%%%%%%%%%%%%%%%%%%%%%%%%%%%%%%%
%%%%%%%%%%%     Eq. 2
${
t_{1,2}, \Delta, V_{1,2} \ll t_0 .
}$
%%%%%%%%%%%%%%%%%%%%%%%%%%%%%%%%%%%%%%%%%%%%%%%%%%%%%%%%%%%%
With the labelling of lattice sites adopted in \eqref{ladder}
the ratio ${{t_2}/{t_1} \equiv \tan \gamma}$,~${0 < \gamma < {\pi}/{4}}$
serves as a measure of kinetic frustration of the model. 
At ${t_2 = 0~(\gamma = 0)}$ and  ${V_2 = 0}$ the model becomes equivalent to 
a unfrustrated  square ladder, whereas the case of equivalent zigzag links
%%%%%%%%%%%%%%%%%%%%%%%%%%%%%%%%%%%%%%%%%%%%%%%%%%%%%%%%%%%
%%%%%%%%%%%     Eq. 3
%%%%%%%%%%%%%%%%%%%%%%%%%%%%%%%%%%%%%%%%%%%%%%%%%%%%%%%%%%%%%%%%%%%%%%%%%%%%%%%%%%%%%%%%%%%%%%%%%%%%
\begin{align}
\label{max-frustr}
	t_1 = t_2 \equiv t ~~(\gamma = {\pi}/{4}), ~~~ V_1 = V_2 \equiv V ,
\end{align}
%%%%%%%%%%%%%%%%%%%%%%%%%%%%%%%%%%%%%%%%%%%%%%%%%%%%%%%%%%%%%%%%%%%%%%%%%%%%%%%%%%%%%%%%%%%%%%%%%%%%
%%%%%%%%%%%%%%%%%%%%%%%%%%%%%%%%%%%%%%%%%%%%%%%%%%%%%%%%%%%%%%
corresponds to full kinetic and dynamical frustration. The  main objective of this work is to 
investigate the low-energy properties of the weakly coupled model~\eqref{ladder} and  trace the 
crossover between  the regimes of a unfrustrated and fully frustrated  ladder, Eq.~\eqref{max-frustr}, 
taking place on varying $\gamma$ and the difference ${V_1 - V_2}$.

The model~\eqref{ladder} does not include n.n.\ interaction between the particles along the chains.
At 1/2-filling, this interaction supports longitudinal CDW correlations~\cite{Haldane-1980};  those would 
overshadow the quantum fluctuations caused by frustration, which is the subject of our primary interest.
The neglect of the longitudinal interaction can be also justified for cold  gases on optical lattices where 
anisotropy of interatomic interaction can be made very strong in experiments with Rydberg-excited atoms~\cite{marcello1,marcello2}.

Using bosonization approach, we derive  the effective field theory to describe universal low-energy properties 
of the system. For a fully frustrated triangular ladder with the condition~\eqref{max-frustr} imposed, the 
collective charge degrees of freedom are gapless even at 1/2 filling, whereas in the spin sector the interplay of frustration, potential bias and interaction leads to dynamical generation of a spectral gap. Most strongly 
fluctuating physical quantities in this regime are local currents circulating around triangular plaquettes 
of the ladder in a sign-alternating way. This is a repulsive version of the Luther-Emery (LE) liquid~\cite{LE}. 
The peculiar property of the LE phase in the present model is the dominance of orbital antiferromagnetic (OAF) 
correlations which  exhibit the slowest power-law spatial decay in the ground state. Pairing correlations also 
display a power-law behavior but are subdominant relative to OAF. 

The issue of ordered states of low-dimensional correlated electron systems with staggered distributions of 
local  currents has a long history. Halperin and Rice~\cite{HR} were the first to realize that the general 
classification of the electron-hole pairing state in solids should necessarily include the possibility of 
orderings with nonzero local charge or spin currents. The search of two-dimensional phases of interacting 
fermions of the OAF type, characterized by a spontaneous breakdown of time reversal symmetry, was boosted by 
the discovery of high-T$_c$ superconductivity in cuprates~\cite{AM, K, NV, Schulz, NJK}. The OAF 
(or $d$-density wave) phases have received considerable attention due to their possible relevance to the 
pseudogap region in the phase diagram of the cuprate superconductors~\cite{varma,chakra}. The possibility of 
the dominance of the OAF correlations or true OAF long-range ordering  in interacting fermionic ladders has been demonstrated in Refs.~\cite{nersesyan-oaf, nlk, FM}.

Interestingly, recent polarized neutron diffraction experiments~\cite{Bounoua_2020,Burges_2021} have demonstrated the existence of a loop-current phase in the two-leg ladder \({\mathrm{Sr}_{14-x} \mathrm{Ca}_x \mathrm{Cu}_{24} \mathrm{O}_{41}}\). The spatial distribution of the observed local
currents follows a single-plaquette pattern: even though parity and time-reversal symmetries are broken, translational invariance is intact. This is in contrast with the OAF phase discussed in this paper, where the currents are spatially staggered, implying the original translational
symmetry is also spontaneously broken.

While  the fully frustrated ladder represents a LE liquid with a finite spin gap and gapless charge, the 
ground state of a square (unfrustrated) ladder is a 1D Mott insulator (MI) with a finite charge gap and
gapless spin. We show in this paper that even a partial frustration breaks the ``spin'' SU(2) symmetry of
the square ladder and generates a finite spin gap, whereas the charge gap vanishes only in the limit of
full dynamical frustration, ${V_1 \to V_2}$. We demonstrate that between the two extremes, the LE  and MI 
phases, there exists a fully gapped phase corresponding to a partially frustrated ladder with a  
spontaneously broken time reversal symmetry and long-ranged OAF order. 

The paper is organized as follows. In Sec.~\ref{noninteract} we discuss the spectral properties of a 
noninteracting triangular ladder. We define the continuum limit of the 1/2-filled model in the original 
(chain) basis and perform the SU(2) transformation of the chiral fermionic fields which brings the Hamiltonian 
to a diagonal form (band basis). In Sec.~\ref{interact-cont} we derive the continuum version of the interaction 
term of the Hamiltonian in both chain and band representations. In Sec.~\ref{bos-model} we discuss the bosonized 
structure of the Hamiltonian for the fully frustrated ladder. With the gapless charge sector described by a 
Gaussian model, we focus our attention on the spin sector of the theory. We describe the onset of a strong-coupling 
regime in the spin sector and the mechanism of generation of the spin gap. In Sec.~\ref{ops} we derive the bosonized 
form of all order parameters relevant to our model. We discuss in much detail the structure of the operators describing local staggered currents on the links of the triangular ladder We prove that for a fully frustrated ladder the OAF correlations follow the slowest algebraic decay as compared to all other correlations in the model. In Sec.~\ref{incomplete} we discuss the properties  of an incompletely frustrated ladder focusing on the limiting cases of weakly frustrated square ladder and almost fully frustrated triangular ladder. We discuss the mechanism of generation of a charge gap and explain 
the onset of OAF long-range order in the system. Our conclusions are compiled in Sec.~\ref{concl}. The paper has an 
Appendix where we provide some details of Abelian bosonization used in the main text.

%%%%%%%%%%%%%%%%%%%%%%%%%%%%%%%%%%%%%%%%%%%%%%%%%%%%%%%%%%%%%%%%%%%%%%%%%%
%%%%%%%   SECTION               NONINTERACTING  LADDER

\section{Noninteracting ladder}\label{noninteract}

The band spectrum of the noninteracting  model  ${(V_1 = V_2 = 0)}$ is given by
%%%%%%%%%%%%%%%%%%%%%%%%%%%%%%%%%%%%%%%%%%%%%%%%%%%%%%%%%%%%%%%%%%%%%%%%%%%%%%%%%%%%%%%%%%%%%%%%%%%%
\begin{align}
\label{band-spec}
	E_{\pm} (k)
	=
	- 2t_0 \cos k \pm \sqrt{\Delta^2 + t^2 _1 + t^2 _2 + 2t_1 t_2 \cos k}
,
\end{align}
%%%%%%%%%%%%%%%%%%%%%%%%%%%%%%%%%%%%%%%%%%%%%%%%%%%%%%%%%%%%%%%%%%%%%%%%%%%%%%%%%%%%%%%%%%%%%%%%%%%%
where ${|k|<\pi}$. In the limit of two decoupled chains and zero bias 
the band ${E_0 (k) = - 2t_0\cos k}$ is doubly degenerate and  1/2-filled.
The particle-hole symmetry of the spectrum, $E_0 (k+\pi) = - E_0 (k)$,
implies that the chemical potential ${\mu=0}$ and the Fermi momenta  ${\pm k_F = \pm \pi/2}$. 
A weak interchain hopping and/or small $\Delta$ cause  splitting of the two bands. In this case
the band spectrum has four Fermi points, $\pm k^+ _F$ and $\pm k^- _F$, which
at the density ${\rho=1}$  satisfy ${k^+ _F + k^- _F = \pi \rho = \pi}$.
In the square  ladder the band splitting is not $k$-dependent, and the spectrum has the nesting 
property~\cite{cnn-1, cnn-2} ${E_+ (k+\pi) = - E_- (k)}$. As a result, the constraint ${\rho=1}$ 
necessarily leads to  ${\mu=0}$. For a triangular ladder, the particle-hole symmetry of the spectrum 
is lost, implying that at 1/2-filling ${\mu \neq 0}$. At ${t_{1,2}, \Delta\ll t_0}$ the splitting of 
the Fermi momenta, $ {k^+ _F -  k^- _F = 2k_0}$ and $\mu$ are estimated as ${k_0 = {h_0}/{v_F}}$ and 
${\mu = t_1 t_2/2t_0}$, where ${v_F = 2t_0 a_0}$ is the Fermi velocity for an isolated 1/2-filled chain 
(here we recovered the lattice spacing along the chains, $a_0$),
%%%%%%%%%%%%%%%%%%%%%%%%%%%%%%%%%%%%%%%%%%%%%%%%%%%%%%%%%%%%%%%%%%%%%%%%%%%%%%%%%%%%%%%%%%%%%%%%%%%%
\begin{align}	
\label{h0-tau0}
	h_0
	=
	\sqrt{\Delta^2 + \tau^2 _0}
\,,\qquad
	\tau_0
	=
	\sqrt{t^2 _1 + t^2 _2}
\,.
\end{align}
%%%%%%%%%%%%%%%%%%%%%%%%%%%%%%%%%%%%%%%%%%%%%%%%%%%%%%%%%%%%%%%%%%%%%%%%%%%%%%%%%%%%%%%%%%%%%%%%%%%%
Contrary to the square ladder,  geometrical frustration makes the 
Fermi velocities in the upper and lower bands different: up to corrections linear in
${k_0 a_0 \ll 1}$,  these  can be estimated  as
%%%%%%%%%%%%%%%%%%%%%%%%%%%%%%%%%%%%%%%%%%%%%%%%%%%%%%%%%%%%%%%%%%%%%%%%%%%%%%%%%%%%%%%%%%%%%%%%%%%%
\begin{align}
\label{Delta-v}
	v_{\pm}
	=
	v_F
	\pm
	\frac{\delta v}{2}
\,,\qquad
	\delta v
	=
	\frac{2t_1 t_2 a_0}{h_0}
\,.
\end{align}
%%%%%%%%%%%%%%%%%%%%%%%%%%%%%%%%%%%%%%%%%%%%%%%%%%%%%%%%%%%%%%%%%%%%%%%%%%%%%%%%%%%%%%%%%%%%%%%%%%%%

A consistent nonperturbative  description of the ladder  at low energies can be achieved by expressing 
the Hamiltonian~\eqref{ladder} and the interaction part in terms of slowly varying chiral (left and right) 
fermionic fields, $R_{\sigma}(x)$ and $L_{\sigma}(x)$, with subsequent bosonization of the whole model.
Since the coupling between the chains is weak, the continuum representation of fermionic lattice operators 
can be defined relative to the limit of two decoupled chains (${k_F = \pi/2}$). Because the site labeling for 
both chains is chosen to be identical  (see Fig.~\ref{fig:zigzag}) we can write
%%%%%%%%%%%%%%%%%%%%%%%%%%%%%%%%%%%%%%%%%%%%%%%%%%%%%%%%%%%%%%%%%%%%%%%%%%%%%%%%%%%%%%%%%%%%%%%
%%%%%%%%%%%%%%%%%%%%%%%%%%%%%%%%%%%%%%%%%%%%%%%%%%%%%%%%%%%%%%%%%%%%%%%%%%%%%%%%%%%%%%%%%%%%%%%%%%%%
\begin{align}
\label{c-cont-limt}
	c_{n\sigma}
	\to
	\sqrt{a_0}
	\left[
		\mathrm{i}^n R_{\sigma} (x) + (-\mathrm{i})^n  L_{\sigma} (x) 
	\right]
,~~
	\sigma = \pm
\,.
\end{align}
%%%%%%%%%%%%%%%%%%%%%%%%%%%%%%%%%%%%%%%%%%%%%%%%%%%%%%%%%%%%%%%%%%%%%%%%%%%%%%%%%%%%%%%%%%%%%%%%%%%%
%%%%%%%%%%%%%%%%%%%%%%%%%%%%%%%%%%%%%%%%%%%%%%%%%%%%%%%%%%%%%%%%%%%%%%%%%%%%%%%%%%%%%%%%%%%%%%%
Passing to the continuum limit and using the correspondence~\eqref{c-cont-limt} we arrive at the expression
displaying a chirally decomposed structure of the noninteracting Hamiltonian density:
%%%%%%%%%%%%%%%%%%%%%%%%%%%%%%%%%%%%%%%%%%%%%%%%%%%%%%%%%%%%%%%%%%%%%%%%%%%%%%%%%%%%%%%%%%%%%%%%%%%%
\begin{align}
\label{zig-fin}
	\mathcal{H}_{0} (x)
	=
	&
	\sum_{\sigma = \pm} 
	\Big[
		R^{\dag}_{\sigma} (x)
		\left(
			-\mathrm{i}
			v_F
			\partial_x
			-
			\sigma
			\Delta
		\right)
		R^{\vphantom{\dag}}_{\sigma} (x)
	+
	L^{\dag}_{\sigma} (x)
		\left(
			\mathrm{i}
			v_F
			\partial_x
			-
			\sigma
			\Delta
		\right)
		L^{\vphantom{\dag}}_{\sigma} (x)
	\Big]
\nonumber
\\
	&
	{-}\tau_0\!
	\left[
		e^{-\mathrm{i}\gamma}
		R^{\dag}_+ (x)
		R^{\vphantom{\dag}}_- (x)
		+
		e^{\mathrm{i}\gamma}
		L^{\dag}_+ (x)
		L^{\vphantom{\dag}}_- (x)
		+
		\mathrm{H.c.}
	\right]
\!.
\end{align}
%%%%%%%%%%%%%%%%%%%%%%%%%%%%%%%%%%%%%%%%%%%%%%%%%%%%%%%%%%%%%%%%%%%%%%%%%%%%%%%%%%%%%%%%%%%%%%%%%%%%
The omitted $\delta v$-term represents a marginal perturbation at the ultraviolet fixed point 
and will be accounted for in Sec.~\ref{bos-model} where the  bosonized structure of the effective 
continuum model is discussed.

To diagonalize $H_0$ we use an SU(2) transformations of the two-component spinors $R$ and $L$:
%%%%%%%%%%%%%%%%%%%%%%%%%%%%%%%%%%%%%%%%%%%%%%%%%%%%%%%%%%%%%%%%%%%%%%%%%%%%%%%%%%%%%%%%%%%%%%%%%%%%
\begin{align}
\label{RL-transform}
	R(x) = \hat{U}_R \mathcal{R}(x)
\,,\qquad
	L(x) = \hat{U}_L \mathcal{L}(x)
\,,
\end{align}
%%%%%%%%%%%%%%%%%%%%%%%%%%%%%%%%%%%%%%%%%%%%%%%%%%%%%%%%%%%%%%%%%%%%%%%%%%%%%%%%%%%%%%%%%%%%%%%%%%%%
where
%%%%%%%%%%%%%%%%%%%%%%%%%%%%%%%%%%%%%%%%%%%%%%%%%%%%%%%%%%%%%%%%%%%%%%%%%%%%%%%%%%%%%%%%%%%%%%%%%%%%
\begin{align}
	R
	=
	\begin{pmatrix}
	R_+ \\ R_-
	\end{pmatrix}
,\quad
	\begin{pmatrix}
	L_+ \\ L_-
	\end{pmatrix}
,
\end{align}
%%%%%%%%%%%%%%%%%%%%%%%%%%%%%%%%%%%%%%%%%%%%%%%%%%%%%%%%%%%%%%%%%%%%%%%%%%%%%%%%%%%%%%%%%%%%%%%%%%%%
with the identical structure for the transformed spinors $\mathcal{R}$ and $\mathcal{L}$.
The SU(2) matrices  $ \hat{U}_R $ and $ \hat{U}_R $ are given by
%%%%%%%%%%%%%%%%%%%%%%%%%%%%%%%%%%%%%%%%%%%%%%%%%%%%%%%%%%%%%%%%%%%%%%%%%%%%%%%%%%%%%%%%%%%%%%%%%%%%
\begin{gather}
\label{UR-L}
	\hat{U}_R = \hat{W}(\gamma)\hat{U} (\beta)
	\,,\quad
	\hat{U}_L = \hat{W}^{\dag}(\gamma)\hat{U} (\beta) 
\,,
\\
	\hat{W}(\gamma) = e^{-\mathrm{i} \gamma \hat{\sigma}_3/2}
\,,
\nonumber
\\
\label{WU}
	\hat{U} (\beta) = e^{-\mathrm{i} \beta \hat{\sigma}_2/2} 
	=
	\begin{pmatrix}
		u, &-v \\
		v, &\hphantom{-}u  
	\end{pmatrix}
\!,
\\
\label{uv-more}
	u^2 - v^2
	=
	\cos \beta
	=
	\frac{\Delta}{h_0}
\,,\quad
	2uv
	=
	\sin \beta
	=
	\frac{\tau_0}{h_0}
\,.
\end{gather}
%%%%%%%%%%%%%%%%%%%%%%%%%%%%%%%%%%%%%%%%%%%%%%%%%%%%%%%%%%%%%%%%%%%%%%%%%%%%%%%%%%%%%%%%%%%%%%%%%%%%
In components the transformations~\eqref{RL-transform} read
%%%%%%%%%%%%%%%%%%%%%%%%%%%%%%%%%%%%%%%%%%%%%%%%%%%%%%%%%%%%%%%%%%%%%%%%%%%%%%%%%%%%%%%%%%%%%%%%%%%%
\begin{align}
	R_{\sigma}
	=
	e^{-\mathrm{i} \sigma \gamma/2}
	\left(
		u
		\mathcal{R}_{\sigma}
		-
		\sigma
		v
		\mathcal{R}_{-\sigma}
	\right)
,
% \nonumber
% \\
\quad
	L_{\sigma}
	=
	e^{\mathrm{i} \sigma \gamma/2}
	\left(
		u
		\mathcal{L}_{\sigma}
		-
		\sigma
		v
		\mathcal{L}_{-\sigma}
	\right)
.
\label{su2-spinor}
\end{align}
%%%%%%%%%%%%%%%%%%%%%%%%%%%%%%%%%%%%%%%%%%%%%%%%%%%%%%%%%%%%%%%%%%%%%%%%%%%%%%%%%%%%%%%%%%%%%%%%%%%%
The new fields $\mathcal{R}_{\sigma}$ and $\mathcal{L}_{\sigma}$ refer to the band representation of the model.
The diagonalized Hamiltonian (with the $\delta v$-term not included) reads
%%%%%%%%%%%%%%%%%%%%%%%%%%%%%%%%%%%%%%%%%%%%%%%%%%%%%%%%%%%%%%%%%%%%%%%%%%%%%%%%%%%%%%%%%%%%%%%%%%%%
\begin{align}
\label{H0-diag}
	\mathcal{H}_0
	=
	\sum_{\sigma}
	\Big[
		\mathcal{R}^{\dag}_{\sigma}
		\left(
			-\mathrm{i}
			v_F
			\partial_x
			-
			\sigma
			h_0
		\right)
		\mathcal{R}^{\vphantom{\dag}}_{\sigma}
	+
		\mathcal{L}^{\dag}_{\sigma}
		\left(
			\mathrm{i}
			v_F
			\partial_x
			-
			\sigma
			h_0
		\right)
		\mathcal{L}^{\vphantom{\dag}}_{\sigma}
	\Big]
.
\end{align}
%%%%%%%%%%%%%%%%%%%%%%%%%%%%%%%%%%%%%%%%%%%%%%%%%%%%%%%%%%%%%%%%%%%%%%%%%%%%%%%%%%%%%%%%%%%%%%%%%%%%
In the band basis, the ``spinful'' 1D fermions are subject to a uniform ``magnetic'' field $h_0$ along the $z$-axis.

\section{Interaction in the continuum limit} \label{interact-cont}

Now we turn to the interchain interaction in \eqref{ladder}. For each chain $\sigma$ we introduce the smooth parts of 
the normal ordered charge and spin  currents describing fluctuations of the corresponding densities:
%%%%%%%%%%%%%%%%%%%%%%%%%%%%%%%%%%%%%%%%%%%%%%%%%%%%%%%%%%%%%%%%%%%%%%%%%%%%%%%%%%%%%%%%%%%%%%%%%%%%
%\begin{subequations}
\begin{align}
%\label{eq:charge-curr_R}
    {J}_{cR} (x) &= \sum_{\sigma} :\!R^{\dag}_{\sigma} (x) R^{\vphantom{\dag}}_{\sigma}(x) \!:
\,,
\nonumber
\\
	{J}_{cL} (x) &= \sum_{\sigma} :\! L^{\dag}_{\sigma} (x) L^{\vphantom{\dag}}_{\sigma}(x) \!:   
\,,
\label{eq:charge-currents}
\\
%%%%%%%%%%%%%%%%%%%%%%%%%%%%%%%%%%%%%%%%%%%%%%%%%%%%%%%%%%%%%%%%%%%%%%%%%%%%%%%%%%%%%%%%%%%%%%%%%%%%
   J^a_{sR} (x) &= \frac{1}{2} \sum_{\sigma\sigma'}  :\! R^{\dag}_{\sigma} (x) \sigma^a _{\sigma\sigma'}R^{\vphantom{\dag}}_{\sigma'}(x) \!:
\,,
\nonumber
\\
   J^a _{sL} (x) &= \frac{1}{2} \sum_{\sigma\sigma'}  :\! L^{\dag}_{\sigma}(x)  \sigma^a _{\sigma\sigma'}L^{\vphantom{\dag}}_{\sigma'}(x) \!: 
\,,
\label{eq:spin-currents}
\end{align}
%\end{subequations}
%%%%%%%%%%%%%%%%%%%%%%%%%%%%%%%%%%%%%%%%%%%%%%%%%%%%%%%%%%%%%%%%%%%%%%%%%%%%%%%%%%%%%%%%%%%%%%%%%%%%
where ${a=x,y,z}$. The  charge and spin currents  satisfy the anomalous U(1) (charge) and SU(2)$_1$ Kac-Moody algebra~\cite{affleck,GNT}, respectively.

The normal ordering prescription in \eqref{eq:charge-currents} and \eqref{eq:spin-currents}, ${:\!A\!: \mathop{=} A - \langle A \rangle_0}$, is 
defined relative to the vacuum of two decoupled chains. As already mentioned in Sec.~\ref{noninteract}, interchain 
hopping  generates a small chemical potential ${\mu \sim \tau^2 _0 /t_0}$ corresponding to 1/2-filling in the 
frustrated model. This fact can be accounted for within a field-theoretical formalism by redefining the normal-ordering prescription for the charge current (the spin currents remain unchanged). Once such normal reordering is assumed,
there is no need to introduce the chemical potential term into the Hamiltonian.

Omitting temporarily the terms responsible for renormalization of the velocities, we arrive at the following charge-spin separated expression for the interaction density 
%%%%%%%%%%%%%%%%%%%%%%%%%%%%%%%%%%%%%%%%%%%%%%%%%%%%%%%%%%%%%%%%%%%%%%%%%%%%%%%%%%%%%%%%%%%%%%%%%%%%
\begin{align}
	\mathcal{H}_{\mathrm{int}}
	&= 
	\mathcal{H}_{\mathrm{int}}^{(c)}
	+
	\mathcal{H}_{\mathrm{int}}^{(s)}
\,,
\nonumber
\\ 
\label{int-charge}
	\mathcal{H}_{\mathrm{int}} ^{(c)}
	&=
	\frac{1}{2}
	g_+
	J_{cR}
	J_{cL}
	+
	g_-
	\mathcal{O}_{\mathrm{umkl}}
\,,
\\
\label{int-spin}
	\mathcal{H}_{\mathrm{int}}^{(s)}
	&=
	-2g_+
	J^z_{sR}
	J^z_{sL}
	-
	g_-
	\mathcal{O}_{\mathrm{back}}
\,.
\end{align}
%%%%%%%%%%%%%%%%%%%%%%%%%%%%%%%%%%%%%%%%%%%%%%%%%%%%%%%%%%%%%%%%%%%%%%%%%%%%%%%%%%%%%%%%%%%%%%%%%%%%
Here
%%%%%%%%%%%%%%%%%%%%%%%%%%%%%%%%%%%%%%%%%%%%%%%%%%%%%%%%%%%%%%%%%%%%%%%%%%%%%%%%%%%%%%%%%%%%%%%%%%%%
\begin{align}
\label{spin-umkl}
	\mathcal{O}_{\mathrm{umkl}} = R^{\dag}_+ R^{\dag}_- L^{\vphantom{\dag}}_- L^{\vphantom{\dag}}_+ + \mathrm{H.c.} \,.
\end{align}
%%%%%%%%%%%%%%%%%%%%%%%%%%%%%%%%%%%%%%%%%%%%%%%%%%%%%%%%%%%%%%%%%%%%%%%%%%%%%%%%%%%%%%%%%%%%%%%%%%%%
is the interchain Umklapp operator and the term
%%%%%%%%%%%%%%%%%%%%%%%%%%%%%%%%%%%%%%%%%%%%%%%%%%%%%%%%%%%%%%%%%%%%%%%%%%%%%%%%%%%%%%%%%%%%%%%%%%%%
\begin{align}
\label{backscat}
	\mathcal{O}_{\mathrm{back}}
	=
	R^{\dag}_+ R^{\vphantom{\dag}}_-  L^{\dag}_- L^{\vphantom{\dag}}_+ + \mathrm{H.c.}
	=
	2 \left(J^x _{sR} J^x _{sL} + J^y _{sR}J^y  _{sL}\right) 
\end{align}
%%%%%%%%%%%%%%%%%%%%%%%%%%%%%%%%%%%%%%%%%%%%%%%%%%%%%%%%%%%%%%%%%%%%%%%%%%%%%%%%%%%%%%%%%%%%%%%%%%%%
describes spin-flip backscattering processes with the momentum transfer ${2k_F = \pi}$.
The coupling constants ${g_{\pm} = g_1 \pm g_2 = (V_1 \pm V_2)a_0}$.
As already mentioned, we assume that interchain  interaction is repulsive: ${g_1,~g_2 > 0}$.
This means that  ${g_+ > 0}$  while the sign of $g_-$ can be arbitrary.

As follows from \eqref{int-charge}, \eqref{int-spin}, the  coupling constant $g_-$ plays a special 
role in the model. In the limit of full dynamical frustration  (${g_- = 0}$) the interchain Umklapp and 
backscattering processes do not contribute to the low-energy dynamics of the system and interaction 
only involves marginal current-current perturbation\footnote{This cancellation is a consequence of 
commensurability of the particle density with the underlying lattice rendering ${2k_F = \pi}$.
It stems from the interplay of position dependent phase factors characterizing the $\pm 2k_F$-amplitudes of the particle densities on different chains. Consider two decoupled and equally populated chains at arbitrary band's filling.
In the continuum limit the density operators are represented as ${\rho_{\sigma} (x) = J_{\sigma} (x) +\tilde{\rho}_{\sigma} (x)},$
where $J_{\sigma} (x) $ are smooth parts of the densities and $\tilde{\rho}_{\sigma} (x)$ are their $\pm 2k_F$-components:
$ {\tilde{\rho}_{\sigma} = e^{-2ik_F x} R^{\dag}_{\sigma} L_{\sigma} + e^{2ik_F x} L^{\dag}_{\sigma} R_{\sigma}}.$
Now consider interaction of the form $
 {V_1 \rho_+ (x) \rho_- (x) +}$ ${V_2 \rho_+ (x) \rho_- (x-a_0)}.
$
The products of the staggered parts of the particle densities acquire the prefactors
$
{V_1 + V_2 e^{\pm 2ik_F a_0}}
$
which at half-filling reduce to ${V_1 - V_2}$. A similar effect of cancellation 
of Umklapp and backscattering perturbations
is known for the $UV$ Hubbard model~\cite{TF}.}
%~\cite{footnote}.
In spite of this fact,
it would be misleading to conclude that at ${g_- = 0}$ the ground state
of the interacting model represents a Tomonaga-Luttinger (TL)  liquid with both charge and spin excitations gapless.
The reason is the fact that due to interchain single-particle hopping the kinetic energy defined in the chain representation, Eq.~\eqref{zig-fin}, is not diagonal in  spin space
and therefore does not commute with the interaction in the spin sector, ${- 2g_+  J^z _{sR}  J^z _{sL}}$.
We will now pass to the band representation in which the kinetic energy is diagonal but interaction, even at ${g_- = 0}$, has a nontrivial, non-Abelian structure which leads to the development of a strong-coupling regime in the spin sector.

In  the rotated basis, the chiral charge and spin currents are defined exactly as in \eqref{eq:charge-currents} and 
\eqref{eq:spin-currents} but with the replacements ${R_{\sigma} \to \mathcal{R}_{\sigma}}$, ${L_{\sigma} \to \mathcal{L}_{\sigma}}$.
Under the SU(2) transformations~\eqref{RL-transform}  the charge currents remain invariant,
while the spin currents in \eqref{int-spin}
undergo SO(3) rotations (for the spin currents we keep the same notations as those in the chain basis)
%%%%%%%%%%%%%%%%%%%%%%%%%%%%%%%%%%%%%%%%%%%%%%%%%%%%%%%%%%%%%%%%%%%%%%%%%%%%%%%%%%%%%%%%%%%%%%%%%%%%
\begin{align}
	 J^x _{sR,L}  &\to
\cos \gamma \left( {J}^x _{sR,L}\cos \beta + {J}^z _{sR,L}\sin \beta\right)
\mp  {J}^y _{sR,L} \sin \gamma\,,
\label{J1-tr}
\\
 J^y _{sR,L}  &\to \pm
\sin\gamma \left( {J}^x _{sR,L}\cos \beta + {J}^z _{sR,L}\sin \beta\right) 
+  {J}^y _{sR,L} \cos \gamma\,,
\label{J2-tr}
\\
 J^z _{sR,L} &\to 
- {J}^x _{sR,L} \sin\beta + {J}^z _{sR,L} \cos\beta  \label{J3-tr}
\,.
\end{align}
%%%%%%%%%%%%%%%%%%%%%%%%%%%%%%%%%%%%%%%%%%%%%%%%%%%%%%%%%%%%%%%%%%%%%%%%%%%%%%%%%%%%%%%%%%%%%%%%%%%%
Using the above relations, we obtain a rather involved general expression for the interaction density in the spin sector
given in the band representation:
%%%%%%%%%%%%%%%%%%%%%%%%%%%%%%%%%%%%%%%%%%%%%%%%%%%%%%%%%%%%%%%%%%%%%%%%%%%%%%%%%%%%%%%%%%%%%%%%%%%%
\begin{align}
\label{spin-int-final}
	\mathcal{H}^{(s)}_{\mathrm{int}}
	=
	&-
	2g_+
	\left(
		J^z_{sR} J^z_{sL}
		\cos^2 \beta
		+
		J^x_{sR} J^x_{sL}
		\sin^2 \beta
	\right)
    +
	g_+
	\left(J^z_{sR} J^x_{sL} + J^x_{sR} J^z_{sL}\right)
	\sin 2\beta
\nonumber
\\
	&-
	2g_-
	\Big\{\!
		\cos 2\gamma
		\Big[
			J^x_{sR} J^x_{sL}
			\cos^2 \beta
			+
			J^z _{sR} J^z _{sL}
			\sin^2 \beta
            +
			\frac{1}{2}
			\left(J^x_{sR} J^z_{sL} + J^z_{sR} J^x_{sL}\right)
			\sin 2\beta
			+
			J^y_{sR} J^y_{sL}
		\Big]
\nonumber
\\
	&\qquad\qquad
    +
	\sin 2\gamma
	\Big[
		\left(J^x_{sR} J^y_{sL} - J^y_{sR} J^x_{sL}\right)
		\cos\beta
        +
		\left(J^z_{sR} J^y_{sL} - J^y_{sR} J^z_{sL}\right)
		\sin\beta
	\Big]
	\Big\}
.
\end{align}
%%%%%%%%%%%%%%%%%%%%%%%%%%%%%%%%%%%%%%%%%%%%%%%%%%%%%%%%%%%%%%%%%%%%%%%%%%%%%%%%%%%%%%%%%%%%%%%%%%%%
For full dynamical frustration, ${g_- = 0}$, ${g_+ = 2g_1}$,  Eq.~\eqref{spin-int-final}  simplifies 
 %%%%%%%%%%%%%%%%%%%%%%%%%%%%%%%%%%%%%%%%%%%%%%%%%%%%%%%%%%%%%%%%%%%%%%%%%%%%%%%%%%%%%%%%%%%%%%%%%%%%
\begin{align}
	 \mathcal{H}^{(s)}_{\mathrm{int}}
 =& - 4g_1 \bigg[\! \left(  \frac{\Delta}{h_0} \right)^2 J^z _{sR} J^z _{sL}
+ \left(  \frac{\tau_0}{h_0} \right)^2 J^x _{sR} J^x _{sL}
 - \left(  \frac{\Delta \tau_0}{h_0 ^2} \right) \left( J^z _{sR} J^x _{sL} + J^x _{sR} J^z _{sL}\right)\!\bigg],
\label{ff-s}
\end{align}
%%%%%%%%%%%%%%%%%%%%%%%%%%%%%%%%%%%%%%%%%%%%%%%%%%%%%%%%%%%%%%%%%%%%%%%%%%%%%%%%%%%%%%%%%%%%%%%%%%%%
 whereas in the charge sector
%%%%%%%%%%%%%%%%%%%%%%%%%%%%%%%%%%%%%%%%%%%%%%%%%%%%%%%%%%%%%%%%%%%%%%%%%%%%%%%%%%%%%%%%%%%%%%%%%%%%
\begin{align}
	\mathcal{H}^{(c)}_{\mathrm{int}} = g_1 J_{cR} J_{cL} \,. \label{ff-c}
\end{align}
%%%%%%%%%%%%%%%%%%%%%%%%%%%%%%%%%%%%%%%%%%%%%%%%%%%%%%%%%%%%%%%%%%%%%%%%%%%%%%%%%%%%%%%%%%%%%%%%%%%%
In the next section we show that the last term in the right-hand side (r.h.s.)\ of \eqref{ff-s} is irrelevant at the ultraviolet 
fixed point. Then the remaining total Hamiltonian in the spin sector coincides with the isospin part of a 
$XZ$-symmetric massless Thirring model in a longitudinal ``magnetic'' field~\cite{JNW}
%%%%%%%%%%%%%%%%%%%%%%%%%%%%%%%%%%%%%%%%%%%%%%%%%%%%%%%%%%%%%%%%%%%%%%%%%%%%%%%%%%%%%%%%%%%%%%%%%%%%
\begin{align}
	\mathcal{H}_s &= \mathcal{H}_{0s} - 2g_{\parallel} J^z _{sR} J^z _{sL} - 2g_{\perp} J^x _{sR} J^x _{sL} - 2h_0 \left( J^z _{sR} + J^z _{sL} \right).
\label{XZ}
\end{align}
%%%%%%%%%%%%%%%%%%%%%%%%%%%%%%%%%%%%%%%%%%%%%%%%%%%%%%%%%%%%%%%%%%%%%%%%%%%%%%%%%%%%%%%%%%%%%%%%%%%%
Here $\mathcal{H}_{0s}$ is the unperturbed Hamiltonian of the spin sector coinciding with the critical SU(2)$_1$-symmetric
Wess-Zumino-Novikov-Witten (WZNW) model~\cite{affleck, affleck2, affl-hald} 
%%%%%%%%%%%%%%%%%%%%%%%%%%%%%%%%%%%%%%%%%%%%%%%%%%%%%%%%%%%%%%%%%%%%%%%%%%%%%%%%%%%%%%%%%%%%%%%%%%%%
\begin{align}
	\mathcal{H}_{0s} = \frac{2\pi v_s}{3} \left( :\!\bm{J}^2_{sR}\!: + :\!\bm{J}^2_{sL}\!:
\right) ,
\label{WZ-ham}
\end{align}
%%%%%%%%%%%%%%%%%%%%%%%%%%%%%%%%%%%%%%%%%%%%%%%%%%%%%%%%%%%%%%%%%%%%%%%%%%%%%%%%%%%%%%%%%%%%%%%%%%%%
and the coupling constants
%%%%%%%%%%%%%%%%%%%%%%%%%%%%%%%%%%%%%%%%%%%%%%%%%%%%%%%%%%%%%%%%%%%%%%%%%%%%%%%%%%%%%%%%%%%%%%%%%%%%
\begin{align}
	g_{\parallel} =  2g_1 \left(\frac{\Delta}{h_0} \right)^2, \qquad
g_{\perp} =  2g_1 \left(\frac{\tau_0}{h_0} \right)^2.
\label{parperp}
\end{align}
%%%%%%%%%%%%%%%%%%%%%%%%%%%%%%%%%%%%%%%%%%%%%%%%%%%%%%%%%%%%%%%%%%%%%%%%%%%%%%%%%%%%%%%%%%%%%%%%%%%%

\section{Fully frustrated ladder: bosonization} \label{bos-model}

We now bosonize  the model of a fully frustrated ladder. Relevant details concerning  the rules of 
Abelian bosonization are collected in Appendix~\ref{app-bos}. For a fully frustrated ladder interaction 
is given by \eqref{ff-s} and \eqref{ff-c}. Note that under the condition ${g_- = 0}$ interaction  does not 
depend on the kinetic frustration parameter $\gamma$. Using the  bosonization rules~\eqref{bos-charge-spin} 
and \eqref{fi-Fi}, first we single out  the charge-spin separated Gaussian part of the model, ${\mathcal{H}_G = \mathcal{H}^{(c)}_G  + \mathcal{H}^{(s)}_G}$
%%%%%%%%%%%%%%%%%%%%%%%%%%%%%%%%%%%%%%%%%%%%%%%%%%%%%%%%%%%%%%%%%%%%%%%%%%%%%%%%%%%%%%%%%%%%%%%%%%%%
\begin{alignat}{2}
	 &\mathcal{H}^{(c)}_{G} =\, &&\frac{v_c}{2} \left[K_c  \left(\partial_x \Theta_c \right)^2 + \frac{1}{K_c} \left(\partial_x \Phi_c\right)^2\right]\!,
 \label{gauss-c}\\
& \mathcal{H}^{(s)}_{G} =\, &&\frac{v_s}{2} \left[K_s  \left(\partial_x \Theta_s \right)^2 + \frac{1}{K_s} \left(\partial_x \Phi_s\right)^2\right]
- \sqrt{\frac{2}{\pi}} h_0 \partial_x \Phi_s,
\label{gauss-s}
\end{alignat}
%%%%%%%%%%%%%%%%%%%%%%%%%%%%%%%%%%%%%%%%%%%%%%%%%%%%%%%%%%%%%%%%%%%%%%%%%%%%%%%%%%%%%%%%%%%%%%%%%%%%
where $v_{c,s}$ are renormalized velocities of gapless collective charge and spin excitations.
For small values of the coupling constant $g_1$ the Luttinger-liquid constants are given by
%%%%%%%%%%%%%%%%%%%%%%%%%%%%%%%%%%%%%%%%%%%%%%%%%%%%%%%%%%%%%%%%%%%%%%%%%%%%%%%%%%%%%%%%%%%%%%%%%%%%
\begin{align}
	K_c
	\simeq 1 - \frac{g_1}{\pi v_c}, \quad
	K_s
	\simeq 1 + \frac{g_{\parallel}}{2\pi v_s}
\,.\label{KcKs}
\end{align}
%%%%%%%%%%%%%%%%%%%%%%%%%%%%%%%%%%%%%%%%%%%%%%%%%%%%%%%%%%%%%%%%%%%%%%%%%%%%%%%%%%%%%%%%%%%%%%%%%%%%
According to \eqref{parperp},  for a fully frustrated ladder it is a
nonzero charge-imbalance $\Delta$ that makes ${g_{\parallel} \neq 0}$ and, hence, 
${K_s > 1}$. This fact will play an important role in the sequel.

The bosonized form of the product $J^x _{sR} J^x _{sL}$ is given by formula~\eqref{xx}.
Up to a multiplicative constant, the mixed-product  operator 
${\tilde{\mathcal{O}}_{zx} ={J}^z _R {J}^x_L + { J}^x _R {J}^z_L}$
appearing in \eqref{ff-s} transforms to 
$ \partial_x \Phi_s \cos \sqrt{2\pi} \Phi_s \cos \sqrt{2\pi} \Theta_s
+ \partial_x \Theta_s \sin \sqrt{2\pi} \Phi_s \sin \sqrt{2\pi} \Theta_s.$
This operator is a conformal scalar and its scaling dimension at the ultraviolet fixed point is 2. Its amplitude
is proportional to $\Delta \tau_0$, so it vanishes when ${\Delta \to 0}$. At ${\Delta \neq 0}$ 
(${K_{s} >1}$)  the scaling dimension of the operator $\tilde{\mathcal{O}}_{zx}$
becomes
$
 {1 +  \left(K_s + {K_s}^{-1} \right) /2 > 2}
$
implying that in an interacting system and at ${\Delta \neq  0}$,  
the $\tilde{\mathcal{O}}_{zx}$-perturbation can be neglected as {irrelevant}. The same conclusion applies to the
second mixed-product operator ${\tilde{\mathcal{O}}_{zy} ={J}^z _R {J}^y_L + 
{ J}^y _R {J}^z_L}$.

Rescaling the charge and spin scalar fields
%%%%%%%%%%%%%%%%%%%%%%%%%%%%%%%%%%%%%%%%%%%%%%%%%%%%%%%%%%%%%%%%%%%%%%%%%%%%%%%%%%%%%%%%%%%%%%%%%%%%
\begin{align}
	\Phi_a \to \sqrt{K_a} \Phi_a, \quad
	\Theta_a \to \Theta_a / \sqrt{K_a}, \quad 
	a=c,s
\label{field-rescal}
\end{align}
%%%%%%%%%%%%%%%%%%%%%%%%%%%%%%%%%%%%%%%%%%%%%%%%%%%%%%%%%%%%%%%%%%%%%%%%%%%%%%%%%%%%%%%%%%%%%%%%%%%%
we arrive at the final expression for the fully bosonized Hamiltonian density of our model in the band representation:
%%%%%%%%%%%%%%%%%%%%%%%%%%%%%%%%%%%%%%%%%%%%%%%%%%%%%%%%%%%%%%%%%%%%%%%%%%%%%%%%%%%%%%%%%%%%%%%%%%%%
\begin{align}
	  \mathcal{H} &= \mathcal{H}_c+ \mathcal{H}_s + \mathcal{H}_{\delta v}\,, \label{total-Hbos}
\\
 \mathcal{H}_c &= \frac{v_c}{2} \left[\left(\partial_x \Theta_c \right)^2 + \left(\partial_x \Phi_c \right)^2  \right]\!,
\label{c-fin}
\\
 \mathcal{H}_s &=\frac{v_s}{2} \left[\left(\partial_x \Theta_s \right)^2 + \left(\partial_x \Phi_s\right)^2\right]
- \sqrt{\frac{2K_s}{\pi}} h_0 \partial_x \Phi_s
\nonumber
\\
&\hphantom{=}
+ \frac{g_{\perp}}{(2\pi\alpha)^2}
\left(  \cos \sqrt{8\pi K_s} \Phi_s + \cos \sqrt{8\pi/K_s} \Theta_s \right)\!,
\label{s-fin}
\\
 \mathcal{H}_{\delta v} &=  \frac{\delta v}{2} \Big[  \frac{1}{\sqrt{K_c K_s}}\partial_x \Theta_c  \partial_x \Theta_s 
+ \sqrt{K_c K_s}\partial_x \Phi_c  \partial_x \Phi_s \Big].
\label{split-v-fin}
\end{align}
%%%%%%%%%%%%%%%%%%%%%%%%%%%%%%%%%%%%%%%%%%%%%%%%%%%%%%%%%%%%%%%%%%%%%%%%%%%%%%%%%%%%%%%%%%%%%%%%%%%%
$ \mathcal{H}_{\delta v} $ in \eqref{split-v-fin} is the bosonized form of the  velocity-splitting term. 
This term, however, plays a minor role. As we show below, the system acquires a finite ``magnetization'' 
${\langle \partial_x \Phi_s\rangle \sim h_0/v_s}$ in the ground state. The term~\eqref{split-v-fin} will therefore 
change the chemical potential by the amount ${\delta v h_0/v_s \sim \tau^2 _0/t_0}$, comparable with its 
original  value. This shift of $\mu$ can be implicitly incorporated into the normal reordering procedure 
for the particle densities.
 
We see that full dynamical frustration ${V_1 = V_2}$ ${(g_-=0)}$ makes the charge sector gapless even at 1/2-filling. 
This sector is described  by the Gaussian model $\mathcal{H}_c$ with interaction dependent  parameter $K_c$. All 
nontrivial correlation effects responsible for the development of a strong-coupling regime, are incorporated
into the spin Hamiltonian~\eqref{s-fin}.

The effective Hamiltonian with the structure~\eqref{s-fin} appears in the model of a 1D Fermi-gas with 
backscattering and spin-nonconserving processes in the presence of an external magnetic field~\cite{GS}, 
and in the theory of two  1D Tomonaga-Luttinger liquids weakly coupled by  single-fermion hopping~\cite{nlk}. 
The important feature of the model~\eqref{s-fin}  is the fact that the $h_0$-term describes the effect of 
an external ``magnetic'' field directed along the $z$-axis, while the sum of the two cosines in \eqref{s-fin}
originates from the current bilinear $J^x _{sR} J^x _{sL}$ which introduces exchange anisotropy in the 
perpendicular ($x$) direction. As a result, all parameters of the Hamiltonian~\eqref{s-fin} are subject to 
renormalization. The RG flow for  this model have been analyzed in detail in Refs.~\cite{GS},~\cite{nlk} using 
a two-cutoff scaling procedure. 

Consider first the case of equally populated chains, ${\Delta=0}$, ${K_s = 1}$. Then, without  the gradient term,
Eq.~\eqref{s-fin} represents the ${\beta^2= 8\pi}$  self-dual  sine-Gordon model~\cite{JKKN, wiegmann,selfdual}
in which the sum of two mutually dual cosines constitutes a strictly marginal perturbation. In this case the 
model~\eqref{s-fin}  occurs in a weak-coupling regime with the self-duality property maintained and the constant 
$g_1$  unrenormalized at all energies.

The tiny balance  between the two cosines gets broken by a finite ``magnetic field'' $h_0$ and interaction 
${g_{\parallel} \sim g_1 (\Delta/h_0)^2}$, both acting in the same direction. Indeed, at ${h_0 \neq 0}$ due to 
nonconservation of the total momentum, the backscattering ``spin-flip'' perturbation
$\cos \sqrt{8\pi K_s} \Phi_s (x)$ gets suppressed at distances ${|x| \sim v_s /h_0 \sqrt{K_s}}$, whereas 
the scaling dimension of the above cosine operator becomes ${2K_s > 2}$ rendering its irrelevance in the RG sense.
At low energies  the effective Hamiltonian transforms to a sine-Gordon model for the dual field $\Theta_s$:
%%%%%%%%%%%%%%%%%%%%%%%%%%%%%%%%%%%%%%%%%%%%%%%%%%%%%%%%%%%%%%%%%%%%%%%%%%%%%%%%%%%%%%%%%%%%%%%%%%%%
\begin{align}
	\mathcal{H}_s =  \mathcal{H}^0 _s
- \sqrt{\frac{2K_s}{\pi}} h_0 \partial_x \Phi_s + \frac{g_{\perp}}{(2\pi\alpha)^2}
\cos \sqrt{8\pi/K_s} \Theta_s \label{s-fin-B} \,.
\end{align}
%%%%%%%%%%%%%%%%%%%%%%%%%%%%%%%%%%%%%%%%%%%%%%%%%%%%%%%%%%%%%%%%%%%%%%%%%%%%%%%%%%%%%%%%%%%%%%%%%%%%
The gradient term in \eqref{s-fin-B} is proportional to the momentum ${\tilde{\Pi}_s = \partial_x \Phi_s}$
conjugate to the dual field $\Theta_s$. Therefore, by Galilean invariance, the system sustains a finite 
``magnetization'' ${\langle \partial_x \Phi_s \rangle \sim h_0}$, that is a finite degree of delocalization of the fermions 
between the chains and a finite charge imbalance. The magnitude of the mass gap in the sine-Gordon model~\eqref{s-fin-B} depends on the ratio $g_{\parallel}/g_{\perp}$~\cite{JNW}, that is on the relation between 
$\Delta$ and $\tau_0$:
%%%%%%%%%%%%%%%%%%%%%%%%%%%%%%%%%%%%%%%%%%%%%%%%%%%%%%%%%%%%%%%%%%%%%%%%%%%%%%%%%%%%%%%%%%%%%%%%%%%%
\begin{alignat}{2}
	m_s
 &\simeq \frac{v_s}{\alpha} \exp \left(-  \frac{\pi^2 v_s}{2g_1}\right), \qquad\qquad\quad&&\Delta \ll \tau_0,
 \label{case1}\\
  &\simeq \frac{v_s}{\alpha} \exp \left( - \frac{\pi v_s}{g_1} \ln \frac{\Delta}{\tau_0} \right),
\qquad &&\Delta \gg \tau_0. \label{case2} 
\end{alignat}
%%%%%%%%%%%%%%%%%%%%%%%%%%%%%%%%%%%%%%%%%%%%%%%%%%%%%%%%%%%%%%%%%%%%%%%%%%%%%%%%%%%%%%%%%%%%%%%%%%%%
In the massive phase the field $\Theta_s$ is locked in one of the degenerate minima of the cosine potential:
%%%%%%%%%%%%%%%%%%%%%%%%%%%%%%%%%%%%%%%%%%%%%%%%%%%%%%%%%%%%%%%%%%%%%%%%%%%%%%%%%%%%%%%%%%%%%%%%%%%%
\begin{align}
	\Theta^{\mathrm{vac}}_n = \sqrt{\frac{\pi K_s}{8}} (1+2n), ~~~n = 0, \pm 1, \pm 2, \ldots 
\label{vac-theta}
\end{align}
%%%%%%%%%%%%%%%%%%%%%%%%%%%%%%%%%%%%%%%%%%%%%%%%%%%%%%%%%%%%%%%%%%%%%%%%%%%%%%%%%%%%%%%%%%%%%%%%%%%%
 
From Eq.~\ref{case2} it follows that at ${\Delta \gg \tau_0}$ the spin gap undergoes a nontrivial renormalization 
and gets gradually suppressed on increasing the ratio $\Delta/\tau_0$.  This physically appealing result indicates 
that, for a fully frustrated ladder, the interchain potential bias tends to drive the system to a completely gapless 
TL regime. 
 
The analysis done in Sec.~\ref{incomplete} shows that the spin sector remains massive and described by the sine-Gordon 
model~\eqref{s-fin-B} even for a partially frustrated ladder. In the latter case the coupling constants $K_s$ and 
$g_{\perp}$ continuously depend on $\gamma$ and $g_-$. The spin sector of the ladder becomes gapless only in the limit 
of unfrustrated  (square) ladder, ${\gamma \to 0}$, ${g_2\to 0}$. On the contrary, an arbitrarily weak frustration makes the charge sector gapped. In a broad range of the parameters characterizing a partially frustrated ladder both the charge and spin sectors are gapped and the ground state displays a long-range OAF  order.

\section{Order parameters}\label{ops}

Order parameters are strongly fluctuating fields of the system represented by particle-hole operators with 
the structure $c^{\dag}_{k\sigma} c^{\vphantom{\dag}}_{k+\pi, \sigma'}$, or pairing operators $c_{k\sigma} c_{-k, \sigma'}$, where 
${k \sim \pm k_F = \pm \pi/2}$. In coordinate space, the order parameters transform to local bilinears of the 
chiral fermionic fields which upon bosonization acquire a multiplicative structure
${\mathcal{O} (x) = \mathcal{O}_c (x)\mathcal{O}_s (x)}$. Here $\mathcal{O}_{c,s}(x)$ are vertex operators defined in the 
charge/spin sectors, each (at a weak interaction) with the Gaussian scaling dimension close to 1/2. With the 
spin sector gapped and the dual filed $\Theta_s$ locked at one of the  values~\eqref{vac-theta}, only the operator 
$\sin \sqrt{2\pi /K_s} \Theta_s$ acquires a nonzero  vacuum expectation value. All other operators are short-ranged 
with ${\langle \mathcal{O}_s \rangle = 0}$. On the other hand, for the fully frustrated ladder (${g_- = 0}$) the charge sector 
is gapless. This means that the operators $\mathcal{O}_c $ will have the structure   
$e^{-\mathrm{i} \sqrt{2\pi K_c} \Phi_c}$ or $e^{-\mathrm{i} \sqrt{2\pi/K_c} \Theta_c}$; accordingly the correlation function 
$\langle \mathcal{O}_c (x) \mathcal{O}_c (0) \rangle$ will decay with the power-law $|x|^{- K_c}$ or  $|x|^{- 1/K_c}$, respectively. 
Since ${K_c < 1}$, dominant correlations in the fully frustrated ladder must be those described by the operator 
%%%%%%%%%%%%%%%%%%%%%%%%%%%%%%%%%%%%%%%%%%%%%%%%%%%%%%%%%%%%%%%%%%%%%%%%%%%%%%%%%%%%%%%%%%%%%%%%%%%%
\begin{align}
	\mathcal{O}_{\mathrm{curr}} \sim e^{-\mathrm{i} \sqrt{2\pi K_c} \Phi_c} \sin \sqrt{2\pi/K_s} \Theta_s \,. \label{curr-op}
\end{align}
%%%%%%%%%%%%%%%%%%%%%%%%%%%%%%%%%%%%%%%%%%%%%%%%%%%%%%%%%%%%%%%%%%%%%%%%%%%%%%%%%%%%%%%%%%%%%%%%%%%%
As shown below, the field~\eqref{curr-op} is a continuum representation of the lattice operators  describing staggered currents on the links of the triangular ladder. The sign-alternating distribution of the circulating local currents along the ladder is a signature of the dominant OAF power-law correlations in the system. Pairing correlations with the order parameter  $\sim$ $e^{-\mathrm{i} \sqrt{2\pi/K_c} \Theta_c}$ also display a power law but are  sub-dominant.

%%%%%%%%%%%%%%%%%%%%%%%%%%%%%%%%%%%%%%%%%%%%%%%%%%%%%%%%%%%%%%%%%%%%%%%%%%%%%%%%%%%%%%%%%%%%%%%%%%%%
%%%%%%%%%%%%%%%          Fig 1  
\begin{figure}[t!]
\centering
\begin{tikzpicture}
%%%%%%%%%%%%%%%%%%%%%%%%%%%%%%%%%%%%%%%%%%%%%%%%%%
% definitions
\newcommand{\nSegments}{4}% = nPoints - 1
\newcommand{\pointDistance}{1.5}% distance between two points on the same subchain
\newcommand{\xOffset}{0.5}% horizontal offset beyond farthest points
\newcommand{\xmin}{-\xOffset}
\pgfmathsetmacro{\xmax}{\pointDistance*\nSegments + \xOffset}
\newcommand{\y}[1]{% y-coordinates of the two subchains
	\ifnum#1=0 0\fi%
	\ifnum#1=1 2\fi%
	\ifnum#1=2 3\fi%
	\ifnum#1=3 5\fi%
}
\newcommand{\point}[2]{({#1 * \pointDistance}, \y{#2})}% (x,y) pair of each point
\newcommand{\pointColor}[1]{%
	\ifnum#1=0 red\fi%
	\ifnum#1=1 blue\fi%
}
\newcommand{\pointSize}{3pt}
\newcommand{\fluxRadius}{0.15*\y{1}}
\newcommand{\indices}[1]{% labels of the indicated points
	\ifnum#1=1 $n-1$\fi%
	\ifnum#1=2 $n\vphantom{n+-1}$\fi%
	\ifnum#1=3 $n+1$\fi%
	\ifnum#1=4 $n+2$\fi%
}
%%%%%%%%%%%%%%%%%%%%%%%%%%%%%%%%%%%%%%%%%%%%%%%%%%%%%%%%%%%%%%%%%%%%%%%%%%%%%%%%%%%%%%%%%%%%%%%%%%%%
%%%%%%%%%%%%%%%          Upper Fig
%%%%%%%%%%%%%%%%%%%%%%%%%%%%%%%%%%%%%%%%%%%%%%%%%%
% straight lines
\foreach \i in {2,3} {%
	\draw (\xmin, \y{\i}) -- (0, \y{\i});%
	\draw (\nSegments*\pointDistance, \y{\i}) -- (\xmax, \y{\i});%
	}
%%%%%%%%%%%%%%%%%%%%%%%%%
\foreach \i in {0,2}
	\draw[postaction=decorate, decoration={markings, mark=at position .6 with {\arrow[scale=1.75]{Stealth}}}] \point{\i}{3} -- \point{\number\numexpr\i+1\relax}{3};
\foreach \i in {1,3}
	\draw[postaction=decorate, decoration={markings, mark=at position .6 with {\arrow[scale=1.75]{Stealth}}}] \point{\number\numexpr\i+1\relax}{3} -- \point{\i}{3};
\foreach \i in {0,2}
	\draw[postaction=decorate, decoration={markings, mark=at position .6 with {\arrow[scale=1.75]{Stealth}}}] \point{\number\numexpr\i+1\relax}{2} -- \point{\i}{2};
\foreach \i in {1,3}
	\draw[postaction=decorate, decoration={markings, mark=at position .6 with {\arrow[scale=1.75]{Stealth}}}] \point{\i}{2} -- \point{\number\numexpr\i+1\relax}{2};
%%%%%%%%%%%%%%%%%%%%%%%%%%%%%%%%%%%%%%%%%%%%%%%%%%
% vertrical lines lines
\foreach \i in {0,2,4}
	\draw[dashed, postaction=decorate, decoration={markings, mark=at position .575 with {\arrow[scale=1.75]{Stealth}}}] \point{\i}{2} -- \point{\i}{3};
\foreach \i in {1,3}
	\draw[dashed, postaction=decorate, decoration={markings, mark=at position .575 with {\arrow[scale=1.75]{Stealth}}}] \point{\i}{3} -- \point{\i}{2};
%%%%%%%%%%%%%%%%%%%%%%%%%%%%%%%%%%%%%%%%%%%%%%%%%%
% zig-zag lines
\foreach \i in {1,3}
	\draw[dashed, postaction=decorate, decoration={markings, mark=at position .575 with {\arrow[scale=1.75]{Stealth}}}] \point{\number\numexpr\i+1\relax}{3} -- \point{\i}{2};
\foreach \i in {0,2}
	\draw[dashed, postaction=decorate, decoration={markings, mark=at position .575 with {\arrow[scale=1.75]{Stealth}}}] \point{\i}{2} -- \point{\number\numexpr\i+1\relax}{3};
%%%%%%%%%%%%%%%%%%%%%%%%%%%%%%%%%%%%%%%%%%%%%%%%%%
\node[right=0.5em] at (\xmax, \y{3}) {$(+)$};
\node[right=0.5em] at (\xmax, \y{2}) {$(-)$};
\node[below=0.5em] at (2*\pointDistance, \y{2}) {$(a)$};
%%%%%%%%%%%%%%%%%%%%%%%%%%%%%%%%%%%%%%%%%%%%%%%%%%
% fluxes
\draw[violet, thick, -{Latex[length=2mm, width=2mm]}] (0.9*\pointDistance, \y{2} + 0.25*\y{1}) arc[start angle=0, end angle=-330, radius=\fluxRadius];
\draw[violet, thick, -{Latex[length=2mm, width=2mm]}] (1.9*\pointDistance - 2*\fluxRadius, \y{2} + 0.25*\y{1}) arc[start angle=-180, end angle=150, radius=\fluxRadius];
\draw[violet, thick, -{Latex[length=2mm, width=2mm]}] (2.9*\pointDistance, \y{2} + 0.25*\y{1}) arc[start angle=0, end angle=-330, radius=\fluxRadius];
\draw[violet, thick, -{Latex[length=2mm, width=2mm]}] (3.9*\pointDistance - 2*\fluxRadius, \y{2} + 0.25*\y{1}) arc[start angle=-180, end angle=150, radius=\fluxRadius];
%%%%%%%%%%%%%%%%%%%%%%%%%%%%%%%%%%%%%%%%%%%%%%%%%%%%%%%%%%%%%%%%%%%%%%%%%%%%%%%%%%%%%%%%%%%%%%%%%%%%
%%%%%%%%%%%%%%%          Lower Fig
%%%%%%%%%%%%%%%%%%%%%%%%%%%%%%%%%%%%%%%%%%%%%%%%%%
% straight lines
\foreach \i in {0,1} {%
	\draw (\xmin, \y{\i}) -- (0, \y{\i});%
	\draw (\nSegments*\pointDistance, \y{\i}) -- (\xmax, \y{\i});%
	}
%%%%%%%%%%%%%%%%%%%%%%%%%
\foreach \i in {0,2}
	\draw[postaction=decorate, decoration={markings, mark=at position .6 with {\arrow[scale=1.75]{Stealth}}}] \point{\i}{1} -- \point{\number\numexpr\i+1\relax}{1};
\foreach \i in {1,3}
	\draw[postaction=decorate, decoration={markings, mark=at position .6 with {\arrow[scale=1.75]{Stealth}}}] \point{\number\numexpr\i+1\relax}{1} -- \point{\i}{1};
\foreach \i in {0,2}
	\draw[postaction=decorate, decoration={markings, mark=at position .6 with {\arrow[scale=1.75]{Stealth}}}] \point{\number\numexpr\i+1\relax}{0} -- \point{\i}{0};
\foreach \i in {1,3}
	\draw[postaction=decorate, decoration={markings, mark=at position .6 with {\arrow[scale=1.75]{Stealth}}}] \point{\i}{0} -- \point{\number\numexpr\i+1\relax}{0};
%%%%%%%%%%%%%%%%%%%%%%%%%%%%%%%%%%%%%%%%%%%%%%%%%%
% vertrical lines lines
\foreach \i in {0,2,4}
	\draw[dashed, postaction=decorate, decoration={markings, mark=at position .575 with {\arrow[scale=1.75]{Stealth}}}] \point{\i}{0} -- \point{\i}{1};
\foreach \i in {1,3}
	\draw[dashed, postaction=decorate, decoration={markings, mark=at position .575 with {\arrow[scale=1.75]{Stealth}}}] \point{\i}{1} -- \point{\i}{0};
%%%%%%%%%%%%%%%%%%%%%%%%%%%%%%%%%%%%%%%%%%%%%%%%%%
% zig-zag lines
\foreach \i in {0,2}
	\draw[dashed, postaction=decorate, decoration={markings, mark=at position .575 with {\arrow[scale=1.75]{Stealth}}}] \point{\number\numexpr\i+1\relax}{1} -- \point{\i}{0};
\foreach \i in {1,3}
	\draw[dashed, postaction=decorate, decoration={markings, mark=at position .575 with {\arrow[scale=1.75]{Stealth}}}] \point{\i}{0} -- \point{\number\numexpr\i+1\relax}{1};
%%%%%%%%%%%%%%%%%%%%%%%%%%%%%%%%%%%%%%%%%%%%%%%%%%
\node[right=0.5em] at (\xmax, \y{1}) {$(+)$};
\node[right=0.5em] at (\xmax, \y{0}) {$(-)$};
\node[below=0.5em] at (2*\pointDistance, \y{0}) {$(b)$};
%%%%%%%%%%%%%%%%%%%%%%%%%%%%%%%%%%%%%%%%%%%%%%%%%%
% fluxes
\draw[violet, thick, -{Latex[length=2mm, width=2mm]}] (0.1*\pointDistance + 2*\fluxRadius, 0.75*\y{1}) arc[start angle=0, end angle=-330, radius=\fluxRadius];
\draw[violet, thick, -{Latex[length=2mm, width=2mm]}] (1.1*\pointDistance, 0.75*\y{1}) arc[start angle=-180, end angle=150, radius=\fluxRadius];
\draw[violet, thick, -{Latex[length=2mm, width=2mm]}] (2.1*\pointDistance + 2*\fluxRadius, 0.75*\y{1}) arc[start angle=0, end angle=-330, radius=\fluxRadius];
\draw[violet, thick, -{Latex[length=2mm, width=2mm]}] (3.1*\pointDistance, 0.75*\y{1}) arc[start angle=-180, end angle=150, radius=\fluxRadius];
%%%%%%%%%%%%%%%%%%%%%%%%%%%%%%%%%%%%%%%%%%%%%%%%%%
\end{tikzpicture}
\caption{Two patterns of OAF ordering on a triangular ladder.}
\label{fig:oafs-ab}
\end{figure}
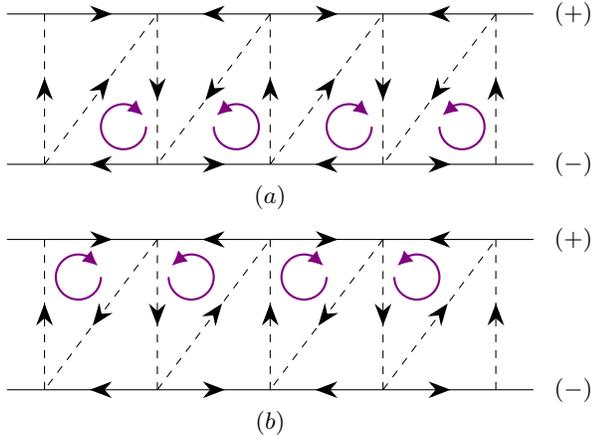
%%%%%%%%%%%%%%%%%%%%%%%%%%%%%%%%%%%%%%%%%%%%%%%%%%%%%%%%%%%%%%%%%%%%%%%%%%%%%%%%%%%%%%%%%%%%%%%%%%%%

In this section we overview the order parameters relevant to our model. We start with the current operators. 
At a semi-classical level, for the OAF patterns shown in Fig.~\ref{fig:oafs-ab}a the currents on the longitudinal 
links, $j^{(\pm)}_0$, and the currents $j_1$ and $j_2$ across the zigzag links should satisfy the relations
%%%%%%%%%%%%%%%%%%%%%%%%%%%%%%%%%%%%%%%%%%%%%%%%%%%%%%%%%%%%%%%%%%%%%%%%%%%%%%%%%%%%%%%%%%%%%%%%%%%%
\begin{align}
	2j^{(+)}_0 = j_1 -  j_2, ~~~~2j^{(-)}_0 = j_1 +  j_2 \,,
\label{Kirch}
\end{align}
%%%%%%%%%%%%%%%%%%%%%%%%%%%%%%%%%%%%%%%%%%%%%%%%%%%%%%%%%%%%%%%%%%%%%%%%%%%%%%%%%%%%%%%%%%%%%%%%%%%%
that follow from Kirchhoff's law. These equations apply to average values of local currents on the corresponding links.
So they will be applicable to the model with a small finite $g_-$ where the charge sector is gapped and the ground state 
has true OAF long-range order. On the other hand, for a fully frustrated ladder (${g_- = 0}$) the charge sector is gapless, 
so all local staggered current operators have zero expectation value. Therefore Eqs.~\eqref{Kirch} can only serve as a guideline when comparing the bosonized definitions of the current operators on horizontal and zigzag links.

First we consider the case of full dynamical frustration, ${g_- = 0}$. The staggered parts of the transverse currents across
the $t_1$ and $t_2$-links are defined as
%%%%%%%%%%%%%%%%%%%%%%%%%%%%%%%%%%%%%%%%%%%%%%%%%%%%%%%%%%%%%%%%%%%%%%%%%%%%%%%%%%%%%%%%%%%%%%%%%%%%
\begin{align}
	&-  \frac{\mathrm{i} t_1}{2} ( c^{\dag}_{n+} c^{\vphantom{\dag}}_{n-} - \mathrm{H.c.} )_{\mathrm{staggered}} 
\to (-1)^n j_1 (x) \,,\label{current1}\\
&   - \frac{\mathrm{i} t_2}{2} ( c^{\dag}_{n+1,+} c^{\vphantom{\dag}}_{n-} - \mathrm{H.c.} )_{\mathrm{staggered}} 
\to (-1)^n j_2 (x) \,,\label{current2}
\end{align}
%%%%%%%%%%%%%%%%%%%%%%%%%%%%%%%%%%%%%%%%%%%%%%%%%%%%%%%%%%%%%%%%%%%%%%%%%%%%%%%%%%%%%%%%%%%%%%%%%%%%
where
%%%%%%%%%%%%%%%%%%%%%%%%%%%%%%%%%%%%%%%%%%%%%%%%%%%%%%%%%%%%%%%%%%%%%%%%%%%%%%%%%%%%%%%%%%%%%%%%%%%%
\begin{align}
	j_1 (x)&=  \frac{t_1 a_0}{2}  \left[  N (x)+ N^{\dag} (x) \right], \nonumber\\
j_2 (x)&=  - \frac{\mathrm{i} t_2 a_0}{2}  \left[  N(x)- N^{\dag}(x) \right],
\end{align}
%%%%%%%%%%%%%%%%%%%%%%%%%%%%%%%%%%%%%%%%%%%%%%%%%%%%%%%%%%%%%%%%%%%%%%%%%%%%%%%%%%%%%%%%%%%%%%%%%%%%
with the fermionic bilinear ${N (x)= R^{\dag} (x)\hat{\sigma}_2 L(x)}$. Due to the
following  property of the matrix $\hat{W}(\gamma)$ in \eqref{WU}, 
${\hat{W}(\gamma)  \hat{\sigma}_2 \hat{W}(\gamma)  = \hat{\sigma}_2}$,
the operator $N$ maintains its form in the band basis as well: ${N (x)= \mathcal{R}^{\dag} (x)\hat{\sigma}_2 \mathcal{L}(x)}$. 
Bosonizing this operator and  rescaling the bosonic fields according to~\eqref{field-rescal} we obtain
%%%%%%%%%%%%%%%%%%%%%%%%%%%%%%%%%%%%%%%%%%%%%%%%%%%%%%%%%%%%%%%%%%%%%%%%%%%%%%%%%%%%%%%%%%%%%%%%%%%%
\begin{align}
j_1 &=  \left(\frac{\tau_0 a}{\pi \alpha} \right)   \cos \gamma  \cos \sqrt{2\pi K_c} \Phi_c \sin \sqrt{2\pi/K_s} \Theta_s,
\nonumber\\
j_2 &=-\left( \frac{\tau_0 a}{\pi \alpha} \right) \sin \gamma  \sin \sqrt{2\pi K_c} \Phi_c\sin \sqrt{2\pi/K_s} \Theta_s.
\label{j1-j2-bos}
\end{align}
%%%%%%%%%%%%%%%%%%%%%%%%%%%%%%%%%%%%%%%%%%%%%%%%%%%%%%%%%%%%%%%%%%%%%%%%%%%%%%%%%%%%%%%%%%%%%%%%%%%%
Being interested in the low-energy (${|E| \ll m_s}$) projections of these operators  onto the gapless charge sector
we can replace the operator $\sin \sqrt{2\pi} \Theta_s$ by  its ground-state expectation value 
%%%%%%%%%%%%%%%%%%%%%%%%%%%%%%%%%%%%%%%%%%%%%%%%%%%%%%%%%%%%%%%%%%%%%%%%%%%%%%%%%%%%%%%%%%%%%%%%%%%%
\begin{align}
	S_s = \langle \sin \sqrt{2\pi/K_s} \Theta_s\rangle_s \sim \left({|m_s|\alpha}/{v_s}\right)^{1/(2K_s)} .
\label{Ss}
\end{align}
%%%%%%%%%%%%%%%%%%%%%%%%%%%%%%%%%%%%%%%%%%%%%%%%%%%%%%%%%%%%%%%%%%%%%%%%%%%%%%%%%%%%%%%%%%%%%%%%%%%%

The staggered part of the longitudinal current on the link $\langle n,n+1 \rangle$ of the $\sigma$-chain is defined as
%%%%%%%%%%%%%%%%%%%%%%%%%%%%%%%%%%%%%%%%%%%%%%%%%%%%%%%%%%%%%%%%%%%%%%%%%%%%%%%%%%%%%%%%%%%%%%%%%%%%
\begin{align}
	- \frac{\mathrm{i} t_0}{2} \left( c^{\dag}_{n\sigma} c^{\vphantom{\dag}}_{n+1,\sigma} - \mathrm{H.c.} \right)_{\mathrm{staggered}}
\to (-1)^n j^{\sigma}_0 (x) \,.
\end{align}
%%%%%%%%%%%%%%%%%%%%%%%%%%%%%%%%%%%%%%%%%%%%%%%%%%%%%%%%%%%%%%%%%%%%%%%%%%%%%%%%%%%%%%%%%%%%%%%%%%%%
Passing to the continuum limit and bosonizing the result, we arrive at the following expression
applicable in the band representation
%%%%%%%%%%%%%%%%%%%%%%%%%%%%%%%%%%%%%%%%%%%%%%%%%%%%%%%%%%%%%%%%%%%%%%%%%%%%%%%%%%%%%%%%%%%%%%%%%%%%
\begin{align}
\label{j0-prop}
	 j_0 ^{(\sigma)} \sim \sigma v_F \left(\frac{\tau_0}{h_0} \right)  \cos  \left( \sqrt{2\pi K_c} \Phi_c  - \sigma \gamma \right)
\partial_x \Phi_s   \sin\sqrt{2\pi/K_s} \Theta_s \,.
\end{align}
%%%%%%%%%%%%%%%%%%%%%%%%%%%%%%%%%%%%%%%%%%%%%%%%%%%%%%%%%%%%%%%%%%%%%%%%%%%%%%%%%%%%%%%%%%%%%%%%%%%%
The exact value of the average $\langle \partial_x \Phi_s   \sin\sqrt{2\pi/K_s} \Theta_s \rangle_s$  for the model~\eqref{s-fin} is unknown.  Clearly, by Galilean invariance ${\partial_x \Phi_s \sim h_0 /v_F}$, so up to a numerical prefactor, the expression~\eqref{j0-prop}  and the current operators~\eqref{j1-j2-bos} are consistent with the Kirchhoff relations~\eqref{Kirch}. 
One may suggest that the longitudinal current operator should be given by
%%%%%%%%%%%%%%%%%%%%%%%%%%%%%%%%%%%%%%%%%%%%%%%%%%%%%%%%%%%%%%%%%%%%%%%%%%%%%%%%%%%%%%%%%%%%%%%%%%%%
\begin{align}
	j_0 ^{(\sigma)} = -\frac{1}{2}\sigma   \left(\frac{\tau_0 a}{\pi \alpha} \right) S_s  
\cos\left( \sqrt{2\pi K_c} \Phi_c - \sigma \gamma  \right) \!.
\label{long-current-final}
\end{align}
%%%%%%%%%%%%%%%%%%%%%%%%%%%%%%%%%%%%%%%%%%%%%%%%%%%%%%%%%%%%%%%%%%%%%%%%%%%%%%%%%%%%%%%%%%%%%%%%%%%%
With this definition and Eqs.~\eqref{j1-j2-bos} the local current operators satisfy the relations~\eqref{Kirch}. 
The scaling dimension  of all  current operators is  $K_c/2$.
 
In the continuum limit, the staggered part of the local density at the $\sigma$-th chain acquires the following bosonized form:
%%%%%%%%%%%%%%%%%%%%%%%%%%%%%%%%%%%%%%%%%%%%%%%%%%%%%%%%%%%%%%%%%%%%%%%%%%%%%%%%%%%%%%%%%%%%%%%%%%%%
\begin{align}
\label{eq:CDW}
	&
	(-1)^n c^{\dag}_{n\sigma} c^{\vphantom{\dag}}_{n\sigma} 
	\to
	R^{\dag}_{\sigma} L^{\vphantom{\dag}}_{\sigma} + \mathrm{H.c.}
	\sim
    e^{-\mathrm{i} \sqrt{2\pi K_c} \Phi_c}
	e^{\mathrm{i} \sigma \gamma}
\nonumber
\\
	&~
	\times
	\bigg[
		\mathrm{i}
		\cos \sqrt{2\pi K_s} \Phi_s
		+
		\sigma
		\Big(\frac{\Delta}{h_0}\Big)
		\sin \sqrt{2\pi K_s} \Phi_s
		+
		\sigma
		\Big(\frac{\tau_0}{h_0}\Big)
		\cos \sqrt{2\pi / K_s} \Theta_s
	\bigg]
	+
	\mathrm{H.c.}
\,.
\end{align}
%%%%%%%%%%%%%%%%%%%%%%%%%%%%%%%%%%%%%%%%%%%%%%%%%%%%%%%%%%%%%%%%%%%%%%%%%%%%%%%%%%%%%%%%%%%%%%%%%%%%
With the vacuum in the spin sector defined by Eq.~\eqref{vac-theta}, the ground state average of the expression in the square brackets of \eqref{eq:CDW} vanishes. Then one concludes that both the total and relative CDW order parameters, defined as \({(-1)^n \sum_{\sigma} c^{\dag}_{n\sigma} c^{\vphantom{\dag}}_{n\sigma}}\) and \({(-1)^n \sum_{\sigma} \sigma c^{\dag}_{n\sigma} c^{\vphantom{\dag}}_{n\sigma}}\), respectively, will have correlation functions exponentially decaying over the distances \({\xi_s \sim v_s / |m_s|}\). The same conclusion is reached for the total and relative in-chain dimerization fields, \({(-1)^n \sum_{\sigma} (c^{\dag}_{n\sigma} c^{\vphantom{\dag}}_{n+1,\sigma} + \mathrm{H.c.})}\) and \({(-1)^n \sum_{\sigma} \sigma (c^{\dag}_{n\sigma} c^{\vphantom{\dag}}_{n+1,\sigma} + \mathrm{H.c.})}\). The transverse BDW operator associated with the $t_1$-link transforms to
%%%%%%%%%%%%%%%%%%%%%%%%%%%%%%%%%%%%%%%%%%%%%%%%%%%%%%%%%%%%%%%%%%%%%%%%%%%%%%%%%%%%%%%%%%%%%%%%%%%%
\begin{align}
\label{eq:BDW-t1}
	(-1)^n
	\sum_{\sigma}
		c^{\dag}_{n\sigma}
		c^{\vphantom{\dag}}_{n,-\sigma}
		\sim
		\cos \sqrt{2\pi K_c} \Phi_c
	\,
    \bigg[
		\Big(\frac{\Delta}{h_0}\Big) 
		\cos \sqrt{\frac{2\pi}{K_s}} \Theta_s 
		-
		\Big(\frac{\tau_0}{h_0}\Big)
		\sin \sqrt{2\pi K_s} \Phi_s
	\bigg].
\end{align}
%%%%%%%%%%%%%%%%%%%%%%%%%%%%%%%%%%%%%%%%%%%%%%%%%%%%%%%%%%%%%%%%%%%%%%%%%%%%%%%%%%%%%%%%%%%%%%%%%%%%
A similar operator related to the $t_2$-link is obtained from~\eqref{eq:BDW-t1} by the replacement \(\cos \sqrt{2\pi K_c} \Phi_c \to \sin \sqrt{2\pi K_c} \Phi_c\). Obviously, both BDW order parameters are short-ranged.

Finally, we turn to pairing correlations. The pairing order parameter associated with the $t_1$-link is defined 
as a smooth part of the  following operator
%%%%%%%%%%%%%%%%%%%%%%%%%%%%%%%%%%%%%%%%%%%%%%%%%%%%%%%%%%%%%%%%%%%%%%%%%%%%%%%%%%%%%%%%%%%%%%%%%%%%
\begin{align}
  \label{S1}
	 &
  c_{n+} c_{n-} \to R_+ L_- + L_+ R_-
 ~\sim  \mathrm{i} e^{-\mathrm{i} \sqrt{2\pi/K_c} \Theta_c} 
\nonumber\\
&\times
 \bigg\{
 \cos\gamma \cos \sqrt{2\pi K_s} \Phi_s
 + \sin\gamma \bigg[ \Big(\frac{\Delta}{h_0}\Big) \sin \sqrt{2\pi K_s} \Phi_s
+\Big(\frac{\tau_0}{h_0}\Big) \cos \sqrt{\frac{2\pi}{K_s}} \Theta_s \bigg]\bigg\}.
\end{align}
%%%%%%%%%%%%%%%%%%%%%%%%%%%%%%%%%%%%%%%%%%%%%%%%%%%%%%%%%%%%%%%%%%%%%%%%%%%%%%%%%%%%%%%%%%%%%%%%%%%%
Both operators in the square brackets of Eq.~\eqref{S1} defined in the spin sector have zero expectation value, implying that the
pairing order parameter $c_{n+} c_{n-}$  is short-ranged. The same conclusion applies to the operator
$ c_{n+} c_{n-1, -} $.

The ``spin-triplet'' longitudinal pairing operators on the two chains with the relative phase $0$ and $\pi$ are defined as
%%%%%%%%%%%%%%%%%%%%%%%%%%%%%%%%%%%%%%%%%%%%%%%%%%%%%%%%%%%%%%%%%%%%%%%%%%%%%%%%%%%%%%%%%%%%%%%%%%%%
\begin{align}
	S^{ 0}_{n,n+1} &=
 \sum_{\sigma} c_{n\sigma} c_{n+1,\sigma}\,, \label{long-0}\\
 S^{\pi}_{n,n+1}&=
 \sum_{\sigma}\sigma  c_{n\sigma} c_{n+1,\sigma}\,.
 \label{long-pi}
\end{align}
%%%%%%%%%%%%%%%%%%%%%%%%%%%%%%%%%%%%%%%%%%%%%%%%%%%%%%%%%%%%%%%%%%%%%%%%%%%%%%%%%%%%%%%%%%%%%%%%%%%%
The operator $S^{ 0}_{n,n+1}$ is invariant under the SU(2) transformations~\eqref{UR-L}.
 Therefore
%%%%%%%%%%%%%%%%%%%%%%%%%%%%%%%%%%%%%%%%%%%%%%%%%%%%%%%%%%%%%%%%%%%%%%%%%%%%%%%%%%%%%%%%%%%%%%%%%%%%
\begin{align}
	S^{ 0}_{n,n+1} &\to - 2\mathrm{i} \sum_{\sigma}R_{\sigma} L_{\sigma} = - 2\mathrm{i} \sum_{\sigma}\mathcal{R}_{\sigma} \mathcal{L}_{\sigma} \sim \mathrm{i}
e^{-\mathrm{i}\sqrt{2\pi/K_c} \Theta_c} \sin \sqrt{2\pi /K_s} \Theta_s\,. \label{0-pair}
\end{align}
%%%%%%%%%%%%%%%%%%%%%%%%%%%%%%%%%%%%%%%%%%%%%%%%%%%%%%%%%%%%%%%%%%%%%%%%%%%%%%%%%%%%%%%%%%%%%%%%%%%%
Furthermore
%%%%%%%%%%%%%%%%%%%%%%%%%%%%%%%%%%%%%%%%%%%%%%%%%%%%%%%%%%%%%%%%%%%%%%%%%%%%%%%%%%%%%%%%%%%%%%%%%%%%
\begin{align}
	 &S^{\pi}_{n,n+1} \to  - 2\mathrm{i} \sum_{\sigma}\sigma R_{\sigma} L_{\sigma}  \sim \mathrm{i} e^{-\mathrm{i}\sqrt{2\pi/K_c} \Theta_c}
 \bigg[
 \Big(\frac{\Delta}{h_0}\Big)  \cos \sqrt{\frac{2\pi}{K_s}} \Theta_s - \Big(\frac{\tau_0}{h_0}\Big) \sin \sqrt{2\pi K_s} \Phi_s
 \bigg].\label{pi-pair}
\end{align}
%%%%%%%%%%%%%%%%%%%%%%%%%%%%%%%%%%%%%%%%%%%%%%%%%%%%%%%%%%%%%%%%%%%%%%%%%%%%%%%%%%%%%%%%%%%%%%%%%%%%
As follows from~\eqref{pi-pair}, $ S^{\pi}$-correlations are short-ranged. On the contrary, at low energies (${|E| \ll m_s}$) the pairing field $S^{0}$ in \eqref{0-pair} can be replaced by 
%%%%%%%%%%%%%%%%%%%%%%%%%%%%%%%%%%%%%%%%%%%%%%%%%%%%%%%%%%%%%%%%%%%%%%%%%%%%%%%%%%%%%%%%%%%%%%%%%%%%
\begin{align}
	S^{ 0} (x) = - \frac{2\mathrm{i}}{\pi\alpha} S_s e^{-\mathrm{i}\sqrt{2\pi/K_c} \Theta_c} \,.
\label{S0-fin}
\end{align}
%%%%%%%%%%%%%%%%%%%%%%%%%%%%%%%%%%%%%%%%%%%%%%%%%%%%%%%%%%%%%%%%%%%%%%%%%%%%%%%%%%%%%%%%%%%%%%%%%%%%
and therefore represents a strongly fluctuating field with a power low correlation
(the quantity $S_s$ is defined in \eqref{Ss}). However, these correlations
fall out with distance as $|x|^{-1/K_c}$ and hence
are sub-dominant relative the OAF ones which decay as $|x|^{-K_c}$.

 \section{OAF long-range order in a partially frustrated ladder}\label{incomplete}
 
In this section we trace the evolution of the properties of the system when passing from the square
  ladder to the fully frustrated triangular ladder by varying the degree of geometrical frustration.
 
\subsection{Square ladder} \label{stanlad}
 
For the square  ladder one sets ${t_2 = g_2 = 0}$, i.e. ${\gamma = 0}$, ${g_{\pm} = g_1}$. In the chain basis, 
the spin part of  interaction given by~\eqref{spin-int-final}  becomes manifestly SU(2)-invariant:
${\mathcal{H}^{(s)}_{\mathrm{int}} =- 2g_1 \bm{J}_{sR} \cdot \bm{J}_{sL}}$. The SU(2) rotation which diagonalizes 
the free part of the Hamiltonian without affecting its interaction part brings the ``magnetic'' terms to a diagonal form,
${\mathcal{H}_{\mathrm{mag}} = - 2 h_0 \left( J^z _{sR} + J^z _{sL} \right)}$. Thus, for the square  ladder, $\mathcal{H}_{\mathrm{int}}$   coincides with the interaction in 1/2-filled  1D Hubbard chain~\cite{affl-hald} in an external magnetic field
%%%%%%%%%%%%%%%%%%%%%%%%%%%%%%%%%%%%%%%%%%%%%%%%%%%%%%%%%%%%%%%%%%%%%%%%%%%%%%%%%%%%%%%%%%%%%%%%%%%%
\begin{align}
	 \mathcal{H}_{\mathrm{int}} =
 \frac{1}{2} g_1 J_{cR} J_{cL} + g_1 \mathcal{O}_{\mathrm{umkl}}
 - 2g_1  \bm{J}_{sR} \cdot \bm{J}_{sL} \,. \label{hub-int}
\end{align}
%%%%%%%%%%%%%%%%%%%%%%%%%%%%%%%%%%%%%%%%%%%%%%%%%%%%%%%%%%%%%%%%%%%%%%%%%%%%%%%%%%%%%%%%%%%%%%%%%%%%
Note that there is an equivalent way to pass to a square ladder: $t_1 = 0$, $g_1 = 0$ 
(i.e.\ $g_+ =- g_- = g_2$, $\gamma = \pi/2$) which leads to the interaction~\eqref{hub-int} with $g_1$ replaced by $g_2$.

The charge excitations of the square ladder are described by the sine-Gordon model
%%%%%%%%%%%%%%%%%%%%%%%%%%%%%%%%%%%%%%%%%%%%%%%%%%%%%%%%%%%%%%%%%%%%%%%%%%%%%%%%%%%%%%%%%%%%%%%%%%%%
\begin{gather}
\label{chargeSG}
	\mathcal{H}_c = \mathcal{H}_c ^0  - \frac{g_1}{2(\pi\alpha)^2} \cos \sqrt{8\pi K_c} \Phi_c \,,
\qquad
	K_c = 1 - \frac{g_1}{2\pi v_c}  < 1 \,,
\end{gather}
%%%%%%%%%%%%%%%%%%%%%%%%%%%%%%%%%%%%%%%%%%%%%%%%%%%%%%%%%%%%%%%%%%%%%%%%%%%%%%%%%%%%%%%%%%%%%%%%%%%%
which is at strong-coupling with a finite commensurability gap $m_c$. The  field $\Phi_c$ is locked at
%%%%%%%%%%%%%%%%%%%%%%%%%%%%%%%%%%%%%%%%%%%%%%%%%%%%%%%%%%%%%%%%%%%%%%%%%%%%%%%%%%%%%%%%%%%%%%%%%%%%
\begin{align}
	\Phi_c^{\mathrm{vac}} = 0 ~~{\mathrm{mod}}~\sqrt{\frac{\pi}{2K_c}} \,. \label{Fi-c-vac}
\end{align}
%%%%%%%%%%%%%%%%%%%%%%%%%%%%%%%%%%%%%%%%%%%%%%%%%%%%%%%%%%%%%%%%%%%%%%%%%%%%%%%%%%%%%%%%%%%%%%%%%%%%
The Hamiltonian in the spin sector is the critical SU(2)$_1$ WZNW model~\eqref{WZ-ham} with a marginally 
irrelevant current-current perturbatioin $ -2g_1 \bm{J}_{sR}\cdot \bm{J}_{sL}$.

From~\eqref{hub-int} it follows that  at $g_1 > 0$ the low-energy structure of the square ladder
is inverted relative to that for the fully frustrated ladder discussed in Sec.~\ref{bos-model}:
the charge sector has a spectral gap, while the spin sector is gapless since the exchange interaction  
is ferromagnetic. Thus the unfrustrated ladder at 1/2-filling displays the properties of a 1D Mott insulator
whose low-energy part of the spectrum coincides with that of a spin-1/2 antiferromagnetic Heisenberg chain.
Note that $\mathcal{H}_{\mathrm{int}}$ does not  depend on $\tau_0$ or $\Delta$. So, for the square ladder,  the effect of the interchain bias $\Delta$ trivially reduces to the replacement of $\tau_0$ by the  effective ``magnetic field'' ${h_0 = \sqrt{\Delta^2 + \tau^2 _0}}$. Since the spin sector is gapless, the response to $h_0$ causes  a finite
``magnetization'' proportional to $h_0$. 

Mapping in the low-energy limit (${|E|\ll m_c}$)  onto the SU(2)$_1$ WZNW  model implies that the strongly 
fluctuating fields are the local fermionic spin density, proportional to the staggered magnetization field 
of the effective antiferromagnetic chain,
%%%%%%%%%%%%%%%%%%%%%%%%%%%%%%%%%%%%%%%%%%%%%%%%%%%%%%%%%%%%%%%%%%%%%%%%%%%%%%%%%%%%%%%%%%%%%%%%%%%%
\begin{align}
\label{N}
	\bm{n}_s 
	\sim 
	\mathrm{Tr} \left(\bm{\sigma} \hat{g}\right)
    \sim \left( \cos \sqrt{2\pi} \Theta_s,\, \sin \sqrt{2\pi} \Theta_s,\,
	-\sin \sqrt{2\pi} \Phi_s \right)\!,
\end{align}
%%%%%%%%%%%%%%%%%%%%%%%%%%%%%%%%%%%%%%%%%%%%%%%%%%%%%%%%%%%%%%%%%%%%%%%%%%%%%%%%%%%%%%%%%%%%%%%%%%%%
and the fermionic bond-density field proportional to the spin dimerization field 
%%%%%%%%%%%%%%%%%%%%%%%%%%%%%%%%%%%%%%%%%%%%%%%%%%%%%%%%%%%%%%%%%%%%%%%%%%%%%%%%%%%%%%%%%%%%%%%%%%%%
\begin{align}
	\epsilon_s \sim \mathrm{Tr} \hat{g} \sim \cos \sqrt{2\pi} \Phi_s \,,
\label{eps}
\end{align}
%%%%%%%%%%%%%%%%%%%%%%%%%%%%%%%%%%%%%%%%%%%%%%%%%%%%%%%%%%%%%%%%%%%%%%%%%%%%%%%%%%%%%%%%%%%%%%%%%%%%
where $\hat{g}$ is the Wess-Zumino matrix field in the  fundamental representatiopn of the SU(2) group~\cite{cft}.
Associated with these spin fields are the staggered parts of the following single-particle operators:
\\
Relative CDW (CDW$^-$)
%%%%%%%%%%%%%%%%%%%%%%%%%%%%%%%%%%%%%%%%%%%%%%%%%%%%%%%%%%%%%%%%%%%%%%%%%%%%%%%%%%%%%%%%%%%%%%%%%%%%
\begin{align}
	(-1)^n {s}^z _n \equiv  \frac{1}{2}(-1)^n \sum_{\sigma} \sigma c^{\dag}_{n\sigma} c^{\vphantom{\dag}}_{n\sigma} \,.
\label{Sz}
\end{align}
%%%%%%%%%%%%%%%%%%%%%%%%%%%%%%%%%%%%%%%%%%%%%%%%%%%%%%%%%%%%%%%%%%%%%%%%%%%%%%%%%%%%%%%%%%%%%%%%%%%%
Transverse BDW (BDW$_{\perp}$)
%%%%%%%%%%%%%%%%%%%%%%%%%%%%%%%%%%%%%%%%%%%%%%%%%%%%%%%%%%%%%%%%%%%%%%%%%%%%%%%%%%%%%%%%%%%%%%%%%%%%
\begin{align}
	(-1)^n{s}^x _n \equiv  \frac{1}{2} (-1)^n\sum_{\sigma} c^{\dag}_{n\sigma} c^{\vphantom{\dag}}_{n, -\sigma} \,. \label{Sx}
\end{align}
%%%%%%%%%%%%%%%%%%%%%%%%%%%%%%%%%%%%%%%%%%%%%%%%%%%%%%%%%%%%%%%%%%%%%%%%%%%%%%%%%%%%%%%%%%%%%%%%%%%%
Transverse staggered current (OAF$_{\perp}$)
%%%%%%%%%%%%%%%%%%%%%%%%%%%%%%%%%%%%%%%%%%%%%%%%%%%%%%%%%%%%%%%%%%%%%%%%%%%%%%%%%%%%%%%%%%%%%%%%%%%%
\begin{align}
	(-1)^n{s}^y _n \equiv  - \frac{\mathrm{i}}{2} (-1)^n\sum_{\sigma} \sigma c^{\dag}_{n\sigma} c^{\vphantom{\dag}}_{n, -\sigma} \,.
\label{Sy}
\end{align}
%%%%%%%%%%%%%%%%%%%%%%%%%%%%%%%%%%%%%%%%%%%%%%%%%%%%%%%%%%%%%%%%%%%%%%%%%%%%%%%%%%%%%%%%%%%%%%%%%%%%
Longitudinal  BDW (BDW$_{\parallel}$): 
%%%%%%%%%%%%%%%%%%%%%%%%%%%%%%%%%%%%%%%%%%%%%%%%%%%%%%%%%%%%%%%%%%%%%%%%%%%%%%%%%%%%%%%%%%%%%%%%%%%%
\begin{align}
	(-1)^n\mathcal{D}_ {n,n+1} \equiv (-1)^n\sum_{\sigma} \left(c^{\dag}_{n\sigma} c^{\vphantom{\dag}}_{n+1, \sigma} + \mathrm{H.c.} \right) .
\label{B}
\end{align}
%%%%%%%%%%%%%%%%%%%%%%%%%%%%%%%%%%%%%%%%%%%%%%%%%%%%%%%%%%%%%%%%%%%%%%%%%%%%%%%%%%%%%%%%%%%%%%%%%%%%
Using the bosonization rules~\eqref{R-L-bos}  and averaging over the massive charge degrees of freedom we obtain the low-energy projections of the order-parameter operators:
%%%%%%%%%%%%%%%%%%%%%%%%%%%%%%%%%%%%%%%%%%%%%%%%%%%%%%%%%%%%%%%%%%%%%%%%%%%%%%%%%%%%%%%%%%%%%%%%%%%%
\begin{align}
\label{bos-n}
\!
	(-1)^n \bm{s}_n= C_c \bm{n}_s (x)
,\quad
	(-1)^n \mathcal{D}_ {n,n+1} =  C_c \epsilon_s (x)
,
\end{align}
%%%%%%%%%%%%%%%%%%%%%%%%%%%%%%%%%%%%%%%%%%%%%%%%%%%%%%%%%%%%%%%%%%%%%%%%%%%%%%%%%%%%%%%%%%%%%%%%%%%%
where 
$
{C_c =  ({a_0}/{\pi\alpha}) \langle \cos \sqrt{2\pi K_c} \Phi_c \rangle_c}
$
and $\bm{n}$  and $\epsilon_s$ are given by \eqref{N} and \eqref{eps}, respectively.
The average  $C_c$ is nonzero   because the charge scalar field is locked at
${\Phi_{c}^{\mathrm{vac}} = 0}$.

Since the spin sector is gapless, none of the  order parameters of the square ladder acquires a nonzero vacuum 
expectation value. Their correlation functions reflect the underlying SU(2) symmetry and 
at distances ${|x| \gg v_c/m_c}$ follow a universal power-law decay ${\sim 1/|x|}$.

\subsection{Weakly frustrated ladder}

Let us find out how the behavior of the square ladder  changes when a weak geometrical frustration is introduced. 
Adding a diagonal link to square plaquettes as shown in Fig.~\ref{fig:zigzag} introduces hopping $t_2$ and interaction 
$V_2$ across this links. Let us first consider the case of a small finite $t_2$ (${\gamma \ll 1}$) with  ${V_2 = 0}$. 
The charge sector remains intact. In the spin Hamiltonian~\eqref{spin-int-final} we keep ${g_+ = g_-=  g_1}$ but take 
into account terms linear and quandratic in the small $\gamma$. The extra terms appearing in the spin Hamiltonian are
%%%%%%%%%%%%%%%%%%%%%%%%%%%%%%%%%%%%%%%%%%%%%%%%%%%%%%%%%%%%%%%%%%%%%%%%%%%%%%%%%%%%%%%%%%%%%%%%%%%%
\begin{align}
	 \Delta \mathcal{H}^{(s)}_{\mathrm{int}} &= 
    4g_1 
    \Big[
     \gamma \cos\beta \left(J^y _{sR} J^x _{sL} - J^x _{sR} J^y _{sL}\right) 
     + 
    \gamma^2 \left(J^x _{sR} J^x _{sL} \cos^2 \beta + J^z _{sR} J^z_{sL} \sin^2 \beta + J^y _{sR} J^y _{sL} \right)
    \Big]
\nonumber\\
     &\hphantom{=}
     +
     O(\gamma^3) \,. 
\label{extra-spin}
\end{align}
%%%%%%%%%%%%%%%%%%%%%%%%%%%%%%%%%%%%%%%%%%%%%%%%%%%%%%%%%%%%%%%%%%%%%%%%%%%%%%%%%%%%%%%%%%%%%%%%%%%%
The perturbation~\eqref{extra-spin} lowers the SU(2) symmetry of the unfrustrated ladder down to XYZ-one.
Recall that for the square ladder  the $J^z_{sR} J^z _{sL}$-term in $\mathcal{H}^{(s)}_{\mathrm{int}}$ is of the form
$
- ({g_1}/{\pi}) \partial_x \varphi_{sR} \partial_x \varphi_{sL},
$
and since ${g_1 > 0}$ the constant ${K_s > 1}$. Small-$\gamma$ corrections cannot change this result.
This means that the terms  $\cos \sqrt{8\pi K_s} \Phi_s$ and $\sin \sqrt{8\pi K_s} \Phi_s$
remain irrelevant.  According to \eqref{xy-yx},  the first line in the r.h.s.\ of \eqref{extra-spin}
reduces to an irrelevant operator
$
{g_1 \gamma} \sin \sqrt{8\pi K_s} \Phi_s.
$
The term $g_1 \gamma^2 J^z_{sR} J^z _{sL}$ in \eqref{extra-spin} induces  a small renormalization of 
$K_s$ keeping the condition ${K_s > 1}$ intact. So $\mathcal{H}^{(s)}_{\mathrm{int}}$ essentially transforms to 
a sine-Gordon model~\eqref{s-fin-B} with the only difference that the amplitude of the cosine operator 
${g_{\perp} = 2g_1 (\tau_0/h_0)^2}$ has to be replaced by a smaller value
%%%%%%%%%%%%%%%%%%%%%%%%%%%%%%%%%%%%%%%%%%%%%%%%%%%%%%%%%%%%%%%%%%%%%%%%%%%%%%%%%%%%%%%%%%%%%%%%%%%%
\begin{align}
	g_{\perp} = 2g_1 \gamma^2 (\tau_0/h_0)^2 \,, \label{g-perp-2}
\end{align}
%%%%%%%%%%%%%%%%%%%%%%%%%%%%%%%%%%%%%%%%%%%%%%%%%%%%%%%%%%%%%%%%%%%%%%%%%%%%%%%%%%%%%%%%%%%%%%%%%%%%
which is second-order effect in kinetic frustration.

Since ${K_s > 1}$, the model flows to strong coupling. We see that introducing a small kinetic frustration 
into the square-ladder model makes the spin sector massive. The total spectrum of such model is \emph{fully} 
gapped. We note that the dual spin field $\Theta_s$ is locked at one of the degenerate vacuum values ~\eqref{vac-theta}. This means that among all operators of the spin sector with scaling dimension close to 1/2 only
$\sin \sqrt{{2\pi }/{K_s}} \Theta_s$ acquires a finite expectation value. The effective spin chain displays a 
Neel long-range order  with ${\langle n_y\rangle \neq 0}$. According to \eqref{Sy} this translates to the OAF long-range
ordering of the system.

Thus, perturbing the square ladder by weak kinetic frustration breaks the spin SU(2) symmetry, opens up a spin gap
and makes the low-energy spectrum fully massive. The effective Hamiltonian of the resulting model is particle-hole
symmetric: it does not contain relevant terms linear in $\gamma$, while the $g_1 \gamma^2$-proportional part
of exchange interaction originates from the particle-hole symmetric term $J^y _{sR} J^y _{sL}$. The order parameters  
CDW$^-$, BDW$_{\perp}$ and BDW$_{\parallel}$ are odd under particle-hole transformation, whereas the OAF order parameter 
is invariant. This explains why a small kinetic frustration selects the OAF fluctuations to make them ``condense'' into a long-range ordered gapped state.

Consider now the distribution of local currents at weak kinetic frustratiion. Since kinetic frustration breaks spin rotational symmetry, the transformation from the chain basis to the band basis becomes nontrivial. The exact 
expressions for the staggered parts of the currents across the $t_1$ and $t_2$ links are given by 
\eqref{j1-j2-bos}. We observe the difference in the charge parts of the two currents (the spin parts for both currents are identical and have a nonzero average). From \eqref{Fi-c-vac} it follows that ${\langle  \sin \sqrt{2\pi K_c} \Phi_c  \rangle = 0}$ implying that no current flows across the $t_2$-diagonal and the interchain current is located of the vertical rungs of the ladder:
%%%%%%%%%%%%%%%%%%%%%%%%%%%%%%%%%%%%%%%%%%%%%%%%%%%%%%%%%%%%%%%%%%%%%%%%%%%%%%%%%%%%%%%%%%%%%%%%%%%%
\begin{align}
	&j_1 =  \left(\frac{\tau_0 a}{\pi \alpha} \right)  \langle \cos \sqrt{2\pi K_c} \Phi_c \rangle_c \,
\langle \sin \sqrt{2\pi/K_s}\Theta_s \rangle _s 
+ O(\gamma^2),\nonumber\\
&j_2 =  0 \,.
\label{currr}
\end{align}
%%%%%%%%%%%%%%%%%%%%%%%%%%%%%%%%%%%%%%%%%%%%%%%%%%%%%%%%%%%%%%%%%%%%%%%%%%%%%%%%%%%%%%%%%%%%%%%%%%%%
 
So, adding a small kinetic frustration generates an OAF long-range order in the system. However, orbital currents
flow around square plaquettes, with the local current on the diagonal links vanishing. Comparing the structure of the bosonized operators $j_1$ and $j_2$ in Eqs.~\eqref{j1-j2-bos} we notice that behind the result~\eqref{currr} is
the  symmetry of the gapped charge sector. The configuration in which the charge field is locked  at ${\Phi_c=0}$
is site-parity ($P_S$) symmetric. The same $P_S$ symmetry is the property of the ordered OAF state on the square
plaquettes of the ladder. The diagonal hopping $t_2$ breaks this symmetry; hence the current $j_2$ must vanish.

It can be shown that the staggered part of the BDW operator associated with the added link $t_2$ also has zero
average because this operator is proportional to $\sin \sqrt{2\pi K_c} \Phi_c$. Thus there is no staggered order 
parameter defined on the frustrating $t_2$-link which has a nonzero expectation value. This link affects the spin 
part of the Hamiltonian by breaking the spin rotational SU(2) symmetry of the model but keeps the nomenclature
of all order parameters originally defined for the unfrustrated ladder.

Now we consider  a weak dynamical frustration of the square ladder: ${V_1 \gg V_2 \neq 0}$, $t_2 = 0 ~(\gamma = 0)$.
From \eqref{spin-int-final} we obtain the spin part of the interaction:
%%%%%%%%%%%%%%%%%%%%%%%%%%%%%%%%%%%%%%%%%%%%%%%%%%%%%%%%%%%%%%%%%%%%%%%%%%%%%%%%%%%%%%%%%%%%%%%%%%%%
\begin{align}
	\mathcal{H}^{(s)}_{\mathrm{int}} = - 2g_1 \bm{J}_{sR} \cdot \bm{J}_{sL}
	+ 2g_2 \cos 2\beta \left(J^x _{sR} J^x _{sL} - J^z _{sR} J^z _{sL}  \right) + 2g_2 J^y _{sR} J^y_{sL}. 
\label{new-spin-A}
\end{align}
%%%%%%%%%%%%%%%%%%%%%%%%%%%%%%%%%%%%%%%%%%%%%%%%%%%%%%%%%%%%%%%%%%%%%%%%%%%%%%%%%%%%%%%%%%%%%%%%%%%%
Since ${g_2 \ll g_1}$, the $zz$-exchange remains ferromagnetic, and still ${K_s > 1}$.
The remaining terms in \eqref{new-spin-A}, proportional to $g_2$, are
%%%%%%%%%%%%%%%%%%%%%%%%%%%%%%%%%%%%%%%%%%%%%%%%%%%%%%%%%%%%%%%%%%%%%%%%%%%%%%%%%%%%%%%%%%%%%%%%%%%%
\begin{align}
	2g_2 \left( \cos 2\beta J^x _{sR} J^x _{sL}  + J^y _{sR} J^y _{sL} \right)
    &= - \frac{g_2}{(2\pi\alpha)^2}
\bigg[\! \cos 2\beta \left(\cos \sqrt{8\pi K_s} \Phi_s + \cos \sqrt{\frac{8\pi}{K_s}} \Theta_s
\right)
\nonumber\\
&\qquad\qquad\qquad\qquad
+ \left(\cos \sqrt{8\pi K_s} \Phi_s - \cos \sqrt{\frac{8\pi}{K_s}} \Theta_s
\right)
\!\bigg].
\end{align}
%%%%%%%%%%%%%%%%%%%%%%%%%%%%%%%%%%%%%%%%%%%%%%%%%%%%%%%%%%%%%%%%%%%%%%%%%%%%%%%%%%%%%%%%%%%%%%%%%%%%
Dropping the irrelevant operator $\cos \sqrt{8\pi K_s} \Phi_s$ we eventually arrive at the sine-Gordon 
model for the dual field $\Theta_s$, Eq.~\eqref{s-fin-B}, in which
%%%%%%%%%%%%%%%%%%%%%%%%%%%%%%%%%%%%%%%%%%%%%%%%%%%%%%%%%%%%%%%%%%%%%%%%%%%%%%%%%%%%%%%%%%%%%%%%%%%%
\begin{align}
	g_{\perp} = 2g_2 (\tau_0/h_0)^2 \label{g-perp-3}.
\end{align}
%%%%%%%%%%%%%%%%%%%%%%%%%%%%%%%%%%%%%%%%%%%%%%%%%%%%%%%%%%%%%%%%%%%%%%%%%%%%%%%%%%%%%%%%%%%%%%%%%%%%
Eq.~\eqref{g-perp-3} differs from \eqref{g-perp-2} by the ${g_1 \gamma^2 \to g_2}$ substitution.

We conclude that a weak dynamical frustration introduced into the square ladder, ${V_2 \ll V_1}$,  ${t_2 = 0}$, has the same effect as a weak kinetic frustration ${t_2 \ll t_1}$,  ${V_2 = 0}$. The mechanism of generating the spin gap with the onset of a long-ranged OAF order is the same in both cases.

\subsection{Almost completely frustrated ladder}
 
Now we discuss the situation when deviations from full geometrical frustration in the triangular ladder are small:
%%%%%%%%%%%%%%%%%%%%%%%%%%%%%%%%%%%%%%%%%%%%%%%%%%%%%%%%%%%%%%%%%%%%%%%%%%%%%%%%%%%%%%%%%%%%%%%%%%%%
\begin{align*}
 0< |g_-| \ll g_+\,,\quad\delta \equiv \frac{\pi}{4} - \gamma \ll 1 \,.
\end{align*}
%%%%%%%%%%%%%%%%%%%%%%%%%%%%%%%%%%%%%%%%%%%%%%%%%%%%%%%%%%%%%%%%%%%%%%%%%%%%%%%%%%%%%%%%%%%%%%%%%%%%
 Up to irrelevant corrections, interaction in the spin sector, Eq.~\eqref{spin-int-final}, becomes 
%%%%%%%%%%%%%%%%%%%%%%%%%%%%%%%%%%%%%%%%%%%%%%%%%%%%%%%%%%%%%%%%%%%%%%%%%%%%%%%%%%%%%%%%%%%%%%%%%%%%
\begin{align}
	\mathcal{H}^{(s)}_{\mathrm{int}} 
    = 
    &
    - 2 (g_+ \cos^2 \beta + 2g_- \delta \sin^2 \beta) J^z _R J^z _L
    - 
    4g_- \delta J^y _R J^y_L  
\nonumber\\
    &
    - 2 (g_+ \sin^2 \beta + 2g_-  \delta \cos^2 \beta) J^x _R J^x _L 
    - 
    2g_- (J^x _{sR} J^y _{sL} - J^y _{sR}J^x _{sL}) \cos \beta.
\label{spin-int-fina-newl}
\end{align}
%%%%%%%%%%%%%%%%%%%%%%%%%%%%%%%%%%%%%%%%%%%%%%%%%%%%%%%%%%%%%%%%%%%%%%%%%%%%%%%%%%%%%%%%%%%%%%%%%%%%
On the one hand, extra perturbations in \eqref{spin-int-fina-newl} lead to small (of the order $g_- \delta$)
renormalization of the parameters appearing in the spin Hamiltonian~\eqref{s-fin}. On the other hand,
the last term in the r.h.s.\ of \eqref{spin-int-fina-newl} adds to the  Hamiltonian~\eqref{s-fin}
a perturbation proportional to $g_-  \sin \sqrt{8\pi K_s} \Phi_s$. Together with the operator $\cos \sqrt{8\pi K_s} \Phi_s$ 
in \eqref{s-fin}, this extra perturbation gets suppressed by the gradient term $h_0 \partial_x \Phi_s$ in the course of renormalization, and, as before, the spin sector will cross over to the regime described by the sine-Gordon model~\eqref{s-fin-B} for the dual field $\Theta_s$. We therefore conclude that small $g_-$ and $\delta$ perturbations 
do not cause qualitative changes in the spin sector in the low-energy limit.

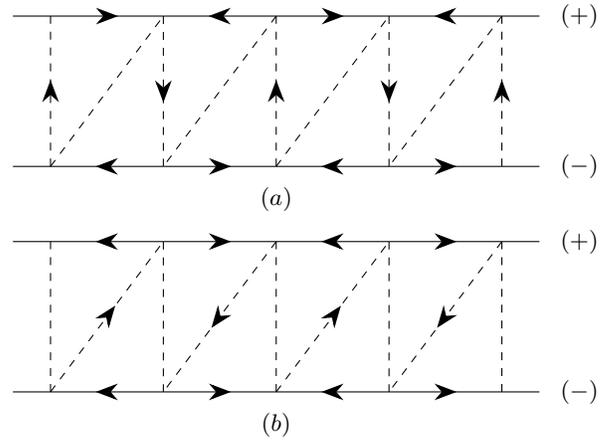
\begin{figure}[t!]
\centering
\begin{tikzpicture}
%%%%%%%%%%%%%%%%%%%%%%%%%%%%%%%%%%%%%%%%%%%%%%%%%%
% definitions
\newcommand{\nSegments}{4}% = nPoints - 1
\newcommand{\pointDistance}{1.5}% distance between two points on the same subchain
\newcommand{\xOffset}{0.5}% horizontal offset beyond farthest points
\newcommand{\xmin}{-\xOffset}
\pgfmathsetmacro{\xmax}{\pointDistance*\nSegments + \xOffset}
\newcommand{\y}[1]{% y-coordinates of the two subchains
	\ifnum#1=0 0\fi%
	\ifnum#1=1 2\fi%
	\ifnum#1=2 3\fi%
	\ifnum#1=3 5\fi%
}
\newcommand{\point}[2]{({#1 * \pointDistance}, \y{#2})}% (x,y) pair of each point
\newcommand{\pointColor}[1]{%
	\ifnum#1=0 red\fi%
	\ifnum#1=1 blue\fi%
}
\newcommand{\pointSize}{3pt}
\newcommand{\fluxRadius}{0.15*\y{1}}
\newcommand{\indices}[1]{% labels of the indicated points
	\ifnum#1=1 $n-1$\fi%
	\ifnum#1=2 $n\vphantom{n+-1}$\fi%
	\ifnum#1=3 $n+1$\fi%
	\ifnum#1=4 $n+2$\fi%
}
%%%%%%%%%%%%%%%%%%%%%%%%%%%%%%%%%%%%%%%%%%%%%%%%%%%%%%%%%%%%%%%%%%%%%%%%%%%%%%%%%%%%%%%%%%%%%%%%%%%%
%%%%%%%%%%%%%%%          Upper Fig
%%%%%%%%%%%%%%%%%%%%%%%%%%%%%%%%%%%%%%%%%%%%%%%%%%
% straight lines
\foreach \i in {2,3} {%
	\draw (\xmin, \y{\i}) -- (0, \y{\i});%
	\draw (\nSegments*\pointDistance, \y{\i}) -- (\xmax, \y{\i});%
	}
%%%%%%%%%%%%%%%%%%%%%%%%%
\foreach \i in {0,2}
	\draw[postaction=decorate, decoration={markings, mark=at position .6 with {\arrow[scale=1.75]{Stealth}}}] \point{\i}{3} -- \point{\number\numexpr\i+1\relax}{3};
\foreach \i in {1,3}
	\draw[postaction=decorate, decoration={markings, mark=at position .6 with {\arrow[scale=1.75]{Stealth}}}] \point{\number\numexpr\i+1\relax}{3} -- \point{\i}{3};
\foreach \i in {0,2}
	\draw[postaction=decorate, decoration={markings, mark=at position .6 with {\arrow[scale=1.75]{Stealth}}}] \point{\number\numexpr\i+1\relax}{2} -- \point{\i}{2};
\foreach \i in {1,3}
	\draw[postaction=decorate, decoration={markings, mark=at position .6 with {\arrow[scale=1.75]{Stealth}}}] \point{\i}{2} -- \point{\number\numexpr\i+1\relax}{2};
%%%%%%%%%%%%%%%%%%%%%%%%%%%%%%%%%%%%%%%%%%%%%%%%%%
% vertrical lines lines
\foreach \i in {0,2,4}
	\draw[dashed, postaction=decorate, decoration={markings, mark=at position .575 with {\arrow[scale=1.75]{Stealth}}}] \point{\i}{2} -- \point{\i}{3};
\foreach \i in {1,3}
	\draw[dashed, postaction=decorate, decoration={markings, mark=at position .575 with {\arrow[scale=1.75]{Stealth}}}] \point{\i}{3} -- \point{\i}{2};
%%%%%%%%%%%%%%%%%%%%%%%%%%%%%%%%%%%%%%%%%%%%%%%%%%
% zig-zag lines
\foreach \i in {1,3}
	\draw[dashed] \point{\number\numexpr\i+1\relax}{3} -- \point{\i}{2};
\foreach \i in {0,2}
	\draw[dashed] \point{\i}{2} -- \point{\number\numexpr\i+1\relax}{3};
%%%%%%%%%%%%%%%%%%%%%%%%%%%%%%%%%%%%%%%%%%%%%%%%%%
\node[right=0.5em] at (\xmax, \y{3}) {$(+)$};
\node[right=0.5em] at (\xmax, \y{2}) {$(-)$};
\node[below=0.5em] at (2*\pointDistance, \y{2}) {$(a)$};
%%%%%%%%%%%%%%%%%%%%%%%%%%%%%%%%%%%%%%%%%%%%%%%%%%%%%%%%%%%%%%%%%%%%%%%%%%%%%%%%%%%%%%%%%%%%%%%%%%%%
%%%%%%%%%%%%%%%          Lower Fig
%%%%%%%%%%%%%%%%%%%%%%%%%%%%%%%%%%%%%%%%%%%%%%%%%%
% straight lines
\foreach \i in {0,1} {%
	\draw (\xmin, \y{\i}) -- (0, \y{\i});%
	\draw (\nSegments*\pointDistance, \y{\i}) -- (\xmax, \y{\i});%
	}
%%%%%%%%%%%%%%%%%%%%%%%%%
\foreach \i in {1,3}
	\draw[postaction=decorate, decoration={markings, mark=at position .6 with {\arrow[scale=1.75]{Stealth}}}] \point{\i}{1} -- \point{\number\numexpr\i+1\relax}{1};
\foreach \i in {0,2}
	\draw[postaction=decorate, decoration={markings, mark=at position .6 with {\arrow[scale=1.75]{Stealth}}}] \point{\number\numexpr\i+1\relax}{1} -- \point{\i}{1};
\foreach \i in {0,2}
	\draw[postaction=decorate, decoration={markings, mark=at position .6 with {\arrow[scale=1.75]{Stealth}}}] \point{\number\numexpr\i+1\relax}{0} -- \point{\i}{0};
\foreach \i in {1,3}
	\draw[postaction=decorate, decoration={markings, mark=at position .6 with {\arrow[scale=1.75]{Stealth}}}] \point{\i}{0} -- \point{\number\numexpr\i+1\relax}{0};
%%%%%%%%%%%%%%%%%%%%%%%%%%%%%%%%%%%%%%%%%%%%%%%%%%
% vertrical lines lines
\foreach \i in {0,2,4}
	\draw[dashed] \point{\i}{0} -- \point{\i}{1};
\foreach \i in {1,3}
	\draw[dashed] \point{\i}{1} -- \point{\i}{0};
%%%%%%%%%%%%%%%%%%%%%%%%%%%%%%%%%%%%%%%%%%%%%%%%%%
% zig-zag lines
\foreach \i in {1,3}
	\draw[dashed, postaction=decorate, decoration={markings, mark=at position .575 with {\arrow[scale=1.75]{Stealth}}}] \point{\number\numexpr\i+1\relax}{1} -- \point{\i}{0};
\foreach \i in {0,2}
	\draw[dashed, postaction=decorate, decoration={markings, mark=at position .575 with {\arrow[scale=1.75]{Stealth}}}] \point{\i}{0} -- \point{\number\numexpr\i+1\relax}{1};
%%%%%%%%%%%%%%%%%%%%%%%%%%%%%%%%%%%%%%%%%%%%%%%%%%
\node[right=0.5em] at (\xmax, \y{1}) {$(+)$};
\node[right=0.5em] at (\xmax, \y{0}) {$(-)$};
\node[below=0.5em] at (2*\pointDistance, \y{0}) {$(b)$};
%%%%%%%%%%%%%%%%%%%%%%%%%%%%%%%%%%%%%%%%%%%%%%%%%%
\end{tikzpicture}
\caption{Patterns of OAF ordering: a) $g_- >0$,  b) $g_- <0$.}
\label{oafs}
\end{figure}

Conversely, as follows from \eqref{int-charge}, a nonzero difference between the coupling constants $V_1$ and $V_2$
introduces a relevant  Umklapp perturbation in the charge sector:
%%%%%%%%%%%%%%%%%%%%%%%%%%%%%%%%%%%%%%%%%%%%%%%%%%%%%%%%%%%%%%%%%%%%%%%%%%%%%%%%%%%%%%%%%%%%%%%%%%%%
\begin{align}
	\mathcal{H}_c &= \frac{v_c}{2}
\left[
\left(\partial_x \Theta_c \right)^2 +
\left(\partial_x \Phi_c \right)^2
\right]  - \frac{g_-}{2(\pi\alpha)^2} \cos \sqrt{8\pi K_c} \Phi_c \,,
\label{charge-new1}
\end{align}
%%%%%%%%%%%%%%%%%%%%%%%%%%%%%%%%%%%%%%%%%%%%%%%%%%%%%%%%%%%%%%%%%%%%%%%%%%%%%%%%%%%%%%%%%%%%%%%%%%%%
with ${K_c < 1}$. The charge sector acquires a mass gap
%%%%%%%%%%%%%%%%%%%%%%%%%%%%%%%%%%%%%%%%%%%%%%%%%%%%%%%%%%%%%%%%%%%%%%%%%%%%%%%%%%%%%%%%%%%%%%%%%%%%
\begin{align}
	m_c \simeq  \frac{v_c}{\alpha} \left(\frac{|g_-|}{\pi v_c}\right)^{\frac{1}{2(1-K_c)}} ,
\end{align}
%%%%%%%%%%%%%%%%%%%%%%%%%%%%%%%%%%%%%%%%%%%%%%%%%%%%%%%%%%%%%%%%%%%%%%%%%%%%%%%%%%%%%%%%%%%%%%%%%%%%
and the field $\Phi_c$ gets locked in one of the degenerate vacua
%%%%%%%%%%%%%%%%%%%%%%%%%%%%%%%%%%%%%%%%%%%%%%%%%%%%%%%%%%%%%%%%%%%%%%%%%%%%%%%%%%%%%%%%%%%%%%%%%%%%
\begin{alignat}{2}
\label{vac>}
	&\Phi_c^{\mathrm{vac}} = 0 \qquad\qquad&&{\mathrm{mod}} ~\sqrt{\frac{\pi}{2K_c}} ~~~{\mathrm{at}}~g_- > 0 
\,,
\\
\label{vac<}
	&\Phi_c^{\mathrm{vac}} =\sqrt{\frac{\pi}{8K_c}} \qquad&&{\mathrm{mod}} ~\sqrt{\frac{\pi}{2K_c}} ~~~{\mathrm{at}}~g_- < 0 \,.
\end{alignat}
%%%%%%%%%%%%%%%%%%%%%%%%%%%%%%%%%%%%%%%%%%%%%%%%%%%%%%%%%%%%%%%%%%%%%%%%%%%%%%%%%%%%%%%%%%%%%%%%%%%%
Thus, at small nonzero $g_-$ both the charge and spin sectors are gapped. 
The immediate consequence of this fact is a spontaneous breakdown of  time reversal symmetry and
the onset of  a true long-range OAF  order in
the system. Indeed, proceeding from the definitions of the  current operators, Eqs.~\eqref{j1-j2-bos}
and \eqref{long-current-final}, we find their expectation values in the ground state (here we consider the
case ${\gamma = \pi/4}$)
%%%%%%%%%%%%%%%%%%%%%%%%%%%%%%%%%%%%%%%%%%%%%%%%%%%%%%%%%%%%%%%%%%%%%%%%%%%%%%%%%%%%%%%%%%%%%%%%%%%%
\begin{align}
	&\langle j_1 \rangle = 2 \langle j_0 \rangle = \frac{1}{\sqrt{2}} \frac{\tau_0 a_0}{\pi\alpha}S_s C_c \,, \nonumber\\
&\langle j_2 \rangle = 0 \,,
\label{posit}
\end{align}
%%%%%%%%%%%%%%%%%%%%%%%%%%%%%%%%%%%%%%%%%%%%%%%%%%%%%%%%%%%%%%%%%%%%%%%%%%%%%%%%%%%%%%%%%%%%%%%%%%%%
at ${g_- > 0}$ and 
%%%%%%%%%%%%%%%%%%%%%%%%%%%%%%%%%%%%%%%%%%%%%%%%%%%%%%%%%%%%%%%%%%%%%%%%%%%%%%%%%%%%%%%%%%%%%%%%%%%%
\begin{align}
	&\langle j_1 \rangle = 0 \,,\nonumber\\
&\langle j_2 \rangle = 2 \langle j_0 \rangle = \frac{1}{\sqrt{2}}
\frac{\tau_0 a_0}{\pi\alpha}S_s S_c \,,
\label{negat}
\end{align}
%%%%%%%%%%%%%%%%%%%%%%%%%%%%%%%%%%%%%%%%%%%%%%%%%%%%%%%%%%%%%%%%%%%%%%%%%%%%%%%%%%%%%%%%%%%%%%%%%%%%
at ${g_- < 0}$. Here $S_s$ is given by Eq.~\eqref{Ss}, while ${C_c = \langle \cos \sqrt{2\pi K_c} \Phi_c \rangle_c}$ and ${S_c = \langle \sin \sqrt{2\pi K_c} \Phi_c \rangle_c}$. The charge averages in \eqref{posit} and \eqref{negat}
scale as $(|m_c| \alpha/v_c)^{K_c/2}$. So, depending on the sign of the difference ${V_1 - V_2}$, we have 
two fully gapped OAF phases shown in Fig.~\ref{oafs}. Each phase is two-fold degenerate with respect 
to sign inversion of local currents on all links. Note that in each of the two cases only one of the 
two transverse links carries a current. Geometrically, these OAF states are those of a square ladder. 
The two realizations in Fig.~\ref{oafs} reflect two ways of passing to the square ladder starting from 
the triangular one. 
 
Thus, the ground-state properties of the weakly frustrated square ladder and  the triangular ladder with 
incomplete dynamical frustration turn out to be qualitatively the same. Continuity arguments give us all 
reasons to claim that at any partial frustration our model features a spontaneous breakdown of time reversal 
symmetry realized as the onset of persistent local currents in the OAF long-range ordered state.

%%%%%%%%%%%%%%%%%%%%%%%%%%%%%%%%%%%%%%%%%%%%%%%%%%%%%%%%%%%%%%%%%%%%%%%%%%%%%%%%%%%%%%%%%%%%%%%%%%%%
%%%%%%%%%%%%%%%          Fig 3
%%%%%%%%%%%%%%%%%%%%%%%%%%%%%%%%%%%%%%%%%%%%%%%%%%%%%%%%%%%%%%%%%%%%%%%%%%%%%%%%%%%%%%%%%%%%%%%%%%%%

\section{Conclusion}\label{concl}

In this paper we have studied the combined effect of geometrical frustration, interchain potential and 
correlations in a model of spinless fermions on a triangular two-chain ladder, Eq.~\eqref{ladder}, at 1/2-filling. 
This model can be regarded as a large-$(t'/t)$ version of the ionic $t$-$t'$ ionic Hubbard chain. On the basis of the derived field-theoretical model we analyzed the low-energy properties of the system in the limit of full kinetic and
dynamical frustration (${t_1 = t_2}$, ${V_1 = V_2}$). Identifying the total particle density with the ``charge'' and the 
relative particle density with the ``spin'' degrees of freedom, we find out that for a fully frustrated ladder the charge sector remains gapless even at 1/2-filling, whereas the spin sector has a spectral gap and is described by a 
$XZ$-symmetric massless Thirring model with a longitudinal ``magnetic'' field. The ground state of the system  in this 
regime represents a repulsive version of the LE liquid. The ``spin'' gap gradually decreases on increasing the ratio 
$\Delta/\tau_0$. The bosonized structure of the order parameters of the model  leads us to the conclusion that at full frustration the dominant correlations exhibiting the slowest power-law decay are those describing staggered orbital 
currants circulating around the triangular plaquettes of the ladder. While the LE phase of the fully frustrated ladder 
and the Mott insulator phase of the square ladder are disordered, the intermediate phase with incomplete dynamical frustration  (${V_1 \neq V_2}$) is fully massive and exhibits long-ranged ordered orbital antiferromagnet state with a spontaneously broken time reversal symmetry. We believe that our findings can be checked experimentally in ultracold 
atom systems on optical lattices.

The present work suggests generalizations in various directions. Firstly, our analysis can be extended to the case of
a geometrically frustrated, charge imbalanced  ladder with  spinful fermions. The field-theoretical description of such model should be realized in terms of the bosonic fields related to the total and relative (``flavor'') charge and total 
and relative (``spin-flavor'') collective degrees of freedom. It is interesting to find out what is the impact of full geometrical frustration and potential bias on the ground state phase diagram of the spinful ladder model. We are 
planning to pursue this research in the future.

The approach developed in this paper can be straightforwardly generalized to frustrated ladders with a larger number of chains. This work naturally leads to a challenging problem of the interplay between OAF order and superconductivity in
a strongly anisotropic triangular lattice with repulsive interaction between the fermions.  For such 2D system a good starting point would be an array of 1D chains with weak kinetic and dynamical couplings along the zigzag links.

\section*{Acknowledgments}

We are grateful to B. Beradze for interesting discussions.

\appendix
\numberwithin{equation}{section}

\section{Details of Abelian bosonization}\label{app-bos}

Here we briefly summarize those facts about Abelian bosonization which are used in the main text of the paper.

The noninteracting part of the Hamiltonian given by Eq.~\eqref{H0-diag} formally describes free massless spinful fermions
with a magnetic field $h_0$ along the $z$-axis. According to the Fermi-Bose equivalence in 1+1 dimensions (see e.g.~\cite{GNT,TG}), the Hamiltonian $\tilde{H}_0$ is equivalent to a sum of two Gaussian models describing collective charge 
and spin excitations in terms of the scalar fields $\Phi_c$, $\Phi_s$ and their duals $\Theta_c$, $\Theta_s$:
%%%%%%%%%%%%%%%%%%%%%%%%%%%%%%%%%%%%%%%%%%%%%%%%%%%%%%%%%%%%%%%%%%%%%%%%%%%%%%%%%%%%%%%%%%%%%%%%%%%%
\begin{align}
	 &
	 \tilde{H}_0 \to \int \mathrm{d}x~\mathcal{H}^0 (x),
	 ~~~
	 \mathcal{H}^0 (x) = \mathcal{H}^0 _c (x) + \mathcal{H}^0 _s (x),
\nonumber\\
	&
	\mathcal{H}^0 _c  = \frac{v_F}{2} \left[  (\partial_x \Phi_c)^2 + (\partial_x \Theta_c)^2\right]\!,
\label{c0}
\\
	&
	\mathcal{H}^0 _s  = \frac{v_F}{2} \left[  (\partial_x \Phi_s)^2 + (\partial_x \Theta_s)^2\right]
- \sqrt{\frac{2}{\pi}} h_0 \partial_x \Phi_s
\,.
\label{s0}
\end{align}
%%%%%%%%%%%%%%%%%%%%%%%%%%%%%%%%%%%%%%%%%%%%%%%%%%%%%%%%%%%%%%%%%%%%%%%%%%%%%%%%%%%%%%%%%%%%%%%%%%%%
To bosonize interaction defined in the band basis, Eqs.~\eqref{int-charge}, \eqref{ff-s}, one needs the bosonic representation for the 
charge and spin currents defined in \eqref{eq:charge-currents} and \eqref{eq:spin-currents}. According to the standard rules
%%%%%%%%%%%%%%%%%%%%%%%%%%%%%%%%%%%%%%%%%%%%%%%%%%%%%%%%%%%%%%%%%%%%%%%%%%%%%%%%%%%%%%%%%%%%%%%%%%%%
\begin{align}
\!\!\!
	{J}_{c\nu} = \sqrt{\frac{2}{\pi}} \partial_x \varphi_{c\nu}, ~~
{J}^z_{s\nu} = \frac{1}{\sqrt{2\pi}} \partial_x \varphi_{s\nu}, ~~
\nu = R,L
\label{bos-charge-spin}
\end{align}
%%%%%%%%%%%%%%%%%%%%%%%%%%%%%%%%%%%%%%%%%%%%%%%%%%%%%%%%%%%%%%%%%%%%%%%%%%%%%%%%%%%%%%%%%%%%%%%%%%%%
with $\varphi_{cR,L}$ and $\varphi_{sR,L}$ are chiral bosonic fields in the respective 
sectors,  defined as
%%%%%%%%%%%%%%%%%%%%%%%%%%%%%%%%%%%%%%%%%%%%%%%%%%%%%%%%%%%%%%%%%%%%%%%%%%%%%%%%%%%%%%%%%%%%%%%%%%%%
\begin{align}
	\varphi_{aR} = \frac{\Phi_a - \Theta_a}{2}, ~\varphi_{aL} = \frac{\Phi_a + \Theta_a}{2},
~(a=c,s).
\label{fi-Fi}
\end{align}
%%%%%%%%%%%%%%%%%%%%%%%%%%%%%%%%%%%%%%%%%%%%%%%%%%%%%%%%%%%%%%%%%%%%%%%%%%%%%%%%%%%%%%%%%%%%%%%%%%%%
Bosonizing the transverse  spin currents requires bosonization of the
fermonic operators $\mathcal{R}_{\sigma}(x)$ and $\mathcal{L}_{\sigma}(x)$. We adopt the following prescription 
%%%%%%%%%%%%%%%%%%%%%%%%%%%%%%%%%%%%%%%%%%%%%%%%%%%%%%%%%%%%%%%%%%%%%%%%%%%%%%%%%%%%%%%%%%%%%%%%%%%%
\begin{align}
\label{R-L-bos}
	\mathcal{R}_{\sigma} (x) &= \frac{\kappa_{\sigma}}{\sqrt{2 \pi \alpha}} e^{\mathrm{i} \pi
\sigma/4} \exp [\mathrm{i} \sqrt{4\pi} {\varphi}_{R\sigma}(x)],
\nonumber\\
\mathcal{L}_{\sigma} (x) &= \frac{\kappa_{\sigma}}{\sqrt{2 \pi \alpha}} e^{\mathrm{i} \pi
\sigma/4} \exp [-\mathrm{i} \sqrt{4\pi} \varphi_{L\sigma}(x)],
\end{align}
%%%%%%%%%%%%%%%%%%%%%%%%%%%%%%%%%%%%%%%%%%%%%%%%%%%%%%%%%%%%%%%%%%%%%%%%%%%%%%%%%%%%%%%%%%%%%%%%%%%%
where
%%%%%%%%%%%%%%%%%%%%%%%%%%%%%%%%%%%%%%%%%%%%%%%%%%%%%%%%%%%%%%%%%%%%%%%%%%%%%%%%%%%%%%%%%%%%%%%%%%%%
\begin{align}
	\varphi_{\nu\sigma} = \frac{\varphi_{c\nu} + \sigma \varphi_{s\nu}}{\sqrt{2}}, ~~~(\nu=R,L),
\end{align}
%%%%%%%%%%%%%%%%%%%%%%%%%%%%%%%%%%%%%%%%%%%%%%%%%%%%%%%%%%%%%%%%%%%%%%%%%%%%%%%%%%%%%%%%%%%%%%%%%%%%
are chiral bosonic fields with a given ``spin'' projection $\sigma$ (it should be kept in mind that here
actually $\sigma$ is the band index). We impose the condition 
$
{\left[\varphi_{R\sigma} (x), \varphi_{L\sigma'}   (x')\right] = ({\mathrm{i}}/{4})\delta_{\sigma\sigma'}}
$,
which guarantees anticommutation of fermionic fields with the same ``spin'' projection.
In \eqref{R-L-bos} the Klein factors $\kappa_{\sigma}$ are introduced to ensure
anticommutation of the fermion fields with opposite spin projections. They
are defined as follows,
${\kappa^{\dag}_{\sigma} = \kappa_{\sigma}}$, ${\{\kappa_+ , \kappa_- \} = 0}$,
 ${\kappa^2 _{\sigma} = 1}$,
and can be represented by the Pauli matrices
$
{\kappa_+ = \bar{\sigma}_1}$, ${\kappa_-= \bar{\sigma}_2}$, 
${\kappa_+ \kappa_- = - \kappa_- \kappa_+ = \mathrm{i} \bar{\sigma}_3}$ .
Since the zero modes $\kappa_{\pm}$ have no dynamics, we can  fix one
of the two eigenvalues of $\bar {\sigma}_3$, e.g. $+1$,  and assume that 
 ${\kappa_+ \kappa_- = - \kappa_- \kappa_+ = \mathrm{i}}$.

The correspondence~\eqref{R-L-bos} implies that
%%%%%%%%%%%%%%%%%%%%%%%%%%%%%%%%%%%%%%%%%%%%%%%%%%%%%%%%%%%%%%%%%%%%%%%%%%%%%%%%%%%%%%%%%%%%%%%%%%%%
\begin{align}
	\mathcal{R}^{\dag}_+ {\mathcal{R}}_-  
&= \frac{1}{2\pi \alpha}e^{-\mathrm{i} \sqrt{8\pi} \varphi_{sR}} \,, \nonumber\\
\mathcal{L}^{\dag}_+ {\mathcal{L}}_-  
&= \frac{1}{2\pi \alpha}e^{\mathrm{i} \sqrt{8\pi} \varphi_{sL}} \,,
\end{align}
%%%%%%%%%%%%%%%%%%%%%%%%%%%%%%%%%%%%%%%%%%%%%%%%%%%%%%%%%%%%%%%%%%%%%%%%%%%%%%%%%%%%%%%%%%%%%%%%%%%%
and leads to the definitions
%%%%%%%%%%%%%%%%%%%%%%%%%%%%%%%%%%%%%%%%%%%%%%%%%%%%%%%%%%%%%%%%%%%%%%%%%%%%%%%%%%%%%%%%%%%%%%%%%%%%
\begin{alignat}{2}
&
J^x _{sR} = \frac{1}{2\pi\alpha} \cos \sqrt{8\pi} \varphi_{sR}, \quad
&&J^y _{sR} = - \frac{1}{2\pi\alpha} \sin\sqrt{8\pi} \varphi_{sR}, \nonumber\\
&
J^x _{sL} = \frac{1}{2\pi\alpha} \cos \sqrt{8\pi} \varphi_{sL}, \quad
&&J^y _{sL} =  \frac{1}{2\pi\alpha} \sin\sqrt{8\pi} \varphi_{sL},
\label{vec-curr-bos}
\end{alignat}
%%%%%%%%%%%%%%%%%%%%%%%%%%%%%%%%%%%%%%%%%%%%%%%%%%%%%%%%%%%%%%%%%%%%%%%%%%%%%%%%%%%%%%%%%%%%%%%%%%%%
which lead to the useful relations
%%%%%%%%%%%%%%%%%%%%%%%%%%%%%%%%%%%%%%%%%%%%%%%%%%%%%%%%%%%%%%%%%%%%%%%%%%%%%%%%%%%%%%%%%%%%%%%%%%%%
\begin{align}
\label{xx}
	J^x_{sR} J^x_{sL}
	&
	=
	-\frac{1}{8(\pi\alpha)^2}
	\!\left(\cos \sqrt{8\pi} \Phi_s + \cos \sqrt{8\pi} \Theta_s\right)
\!,
\\
\label{yy}
	J^y_{sR} J^y _{sL}
	&
	=
	-\frac{1}{8(\pi\alpha)^2}
	\!\left(\cos \sqrt{8\pi} \Phi_s - \cos \sqrt{8\pi} \Theta_s\right)
\!,
\\
\label{xy+yx}
 	J^x_{sR} J^y_{sL}
 	+
 	J^y_{sR} J^x _{sL}
 	&
	=
	-\frac{1}{(2\pi\alpha)^2}
	\sin \sqrt{8\pi} \Theta_s
\,,
\\
\label{xy-yx}
	J^x_{sR} J^y_{sL}
	-
	J^y_{sR} J^x_{sL}
	&
	=
	-\frac{1}{(2\pi\alpha)^2}
	\sin \sqrt{8\pi} \Phi_s
\,.
\end{align}
%%%%%%%%%%%%%%%%%%%%%%%%%%%%%%%%%%%%%%%%%%%%%%%%%%%%%%%%%%%%%%%%%%%%%%%%%%%%%%%%%%%%%%%%%%%%%%%%%%%%
The velocity-splitting operator is bosonized as follows
%%%%%%%%%%%%%%%%%%%%%%%%%%%%%%%%%%%%%%%%%%%%%%%%%%%%%%%%%%%%%%%%%%%%%%%%%%%%%%%%%%%%%%%%%%%%%%%%%%%%
\begin{align}
	-\mathrm{i} \sum_{\sigma} \sigma \left(\mathcal{R}^{\dag}_{\sigma} \partial_x \mathcal{R} _{\sigma} -  \mathcal{L}^{\dag}_{\sigma} \partial_x \mathcal{L} _{\sigma} \right)
    &= 
    \sum_{\sigma} \sigma \left[ \left( \partial_x \varphi_{R\sigma}\right)^2 +    \left( \partial_x \varphi_{L\sigma}\right)^2\right]
\nonumber\\
    &= \left(\partial_x \Phi_c \partial_x \Phi_s + \partial_x \Theta_c \partial_x \Theta_s 
\right). \label{delta-v-bos}
\end{align}
%%%%%%%%%%%%%%%%%%%%%%%%%%%%%%%%%%%%%%%%%%%%%%%%%%%%%%%%%%%%%%%%%%%%%%%%%%%%%%%%%%%%%%%%%%%%%%%%%%%%

%%%%%%%%%%%%%%%%%%%%%%%%%%%%%%%%%%%%%%%%%%%%%%%%%%%%%%%%%%%%%%%%%%%%%%%%%%
%%%%%%%   SECTION                    REFERENCES

\end{document}